\documentclass{aa}  
\usepackage[stable]{footmisc}
\usepackage{subfigure}
\usepackage{graphicx}
\usepackage{natbib}
\usepackage{txfonts}
\usepackage{ulem}
\begin{document}
 \title{The molecular distribution of the IRDC $\rm{G}351.77-0.51$}
 \author{S. Leurini
        \inst{1}
        \and T. Pillai \inst{2,3}
        \and T. Stanke \inst{4}   \and F. Wyrowski \inst{1} \and L. Testi \inst{4,5} \and F. Schuller \inst{1} \and K. M. Menten \inst{1} \and S. Thorwirth \inst{6}
        }

 \offprints{S. Leurini}

\institute{Max-Planck-Institut f\"ur Radioastronomie, Auf dem H\"ugel 69, 53121 Bonn, Germany\\
\email{sleurini@mpifr.de}
       \and Caltech, 1200 E. California Blvd, Pasadena, CA 91125, USA
       \and Harvard-Smithsonian Center for Astrophysics, 60 Garden Street, Cambridge, MA 02138, USA
       \and ESO, Karl-Schwarzschild Strasse 2, 85748 Garching-bei-M\"unchen, Germany
       \and INAF - Osservatorio Astrofisico di Arcetri, Largo Fermi 5, 50125 Firenze, Italy
       \and I. Physikalisches Institut, Universit\"at zu K\"oln, Z\"ulpicher Str. 77, 50937 K\"oln, Germany
           }

 \date{\today}

\abstract
 {Infrared dark clouds are massive, dense clouds seen in extinction against the IR Galactic background.
Many of these objects appear to be on the verge of star and star cluster formation.}
 {Our aim is to understand the physical properties of IRDCs in very early evolutionary phases. We selected the filamentary IRDC $\rm{G}351.77-0.51$, which is remarkably IR quiet at  8$\mu$m.}
 {As a first step, we  observed mm dust continuum emission and rotational lines of moderate and dense gas
tracers to characterise different condensations along the IRDC and study the velocity field of the filament.}
 {Our initial study confirms coherent velocity
distribution along the infrared dark cloud ruling out any coincidental projection effects. Excellent correlation between MIR
  extinction, mm continuum emission and gas distribution is found. %
 Large-scale turbulence and line profiles throughout the filament is indicative of a shock in this cloud. 
Excellent correlation between line width, and MIR brightness indicates turbulence driven by local star formation.
}
{}


 \maketitle
%

\section{Introduction}\label{intro}
Although the majority of stars, especially massive stars, are observed to form in clusters
\citep{1997A&A...320..159T,1999A&A...342..515T,2003ARA&A..41...57L,2004A&A...425..937D}, our understanding of star formation is
mainly based on studies of nearby star-forming regions  
that mostly give birth to isolated low-mass stars. On the other hand, little is known about the formation of clusters
and of stars in clusters.
Naturally, studies of clusters should begin by
understanding their youngest evolutionary phases.  However, the
youngest (proto)clusters would still be deeply embedded within their natal
dense clouds, thus obscured from view except at far IR/millimetre
wavelengths. Due to the  coarse angular resolution of millimetre observations,
studies of deeply embedded  (proto)clusters have so far focused on the nearest star
forming regions in Ophiuchus \citep{1998A&A...336..150M,2004ApJ...611L..45J,2007A&A...472..519A},
Perseus \citep{2010ApJ...711..655J} and Serpens \citep{1998ApJ...508L..91T,2000ApJ...540L..53T,2002A&A...392.1053O}.
These regions, however, do not form massive stars and rich clusters and it is therefore necessary to extend
these studies to more distant massive star-forming regions. Observational studies of the initial conditions for the
formation of massive clusters have so far been very limited \citep[e.g.][]{1998ApJ...505L..39M,2002ApJ...570..758M,2004science} but
will soon find a new renaissance
with the high angular resolution and sensitivity offered by  the Atacama Large Millimeter/submillimeter Array (ALMA).

Infrared dark clouds (IRDCs) are ideal candidates for studying the earliest
phases of the formation of clusters. They were
discovered as dark patches in the sky at IR wavelengths against
the bright Galactic background during mid-IR imaging surveys with the Infrared Space Observatory \citep[ISO,][]{1996A&A...315L.165P} and the Mid-course Space Experiment \citep[MSX,][]{1998ApJ...494L.199E}.
Their spatial coincidence with
emission from molecular lines and dust indicates that they consist of dense molecular gas and dust of a few
$10^5$~cm$^{-3}$ density \citep{1998ApJ...508..721C,2000ApJ...543L.157C}.  Temperatures in these
clouds are low enough \citep[10--20 K,][]{2006A&A...450..569P} that they do not radiate
significantly even at mid-IR wavelengths: of the 190 dust cores in IRDCs surveyed by \citet{2010ApJ...715..310R}
only 93 have detectable emission at 24$\mu$m. 

The formation mechanism of IRDC is still a matter of debate \citep{2009ApJ...700.1609M}. Recent results 
\citep[e.g., ][]{2010MNRAS.406..187J,2011ApJ...730...44H} 
may  support models of filament formation from converging flows. However, there is still no solid conclusion about 
their origin.
\citet{2007IAUS..237..258T} suggests that IRDC may contain 
the sites of future massive star and star cluster formation, since their densities 
and mass surface densities  are similar to regions
known to be undergoing such formation activity.
 Indeed, emerging evidence suggests
that the earliest phases of the formation of high-mass star and cluster  
are occurring within these clouds  \citep[e.g., ][]{2006A&A...447..929P,2006ApJ...641..389R}.
Studies of IRDCs have often focused on sources that already have
intense star formation activity and are therefore  in late protostellar
evolutionary phases \citep[e.g., ][]{2006A&A...447..929P,2008ApJ...689.1141R}.  
However, to address questions related to   the formation of  
clusters, and of the single stars in them, studies of   earlier evolutionary
phases with no or modest star formation are needed.

\begin{figure*}
\centering
\subfigure[][]{
\includegraphics[angle=-90, width=9cm]{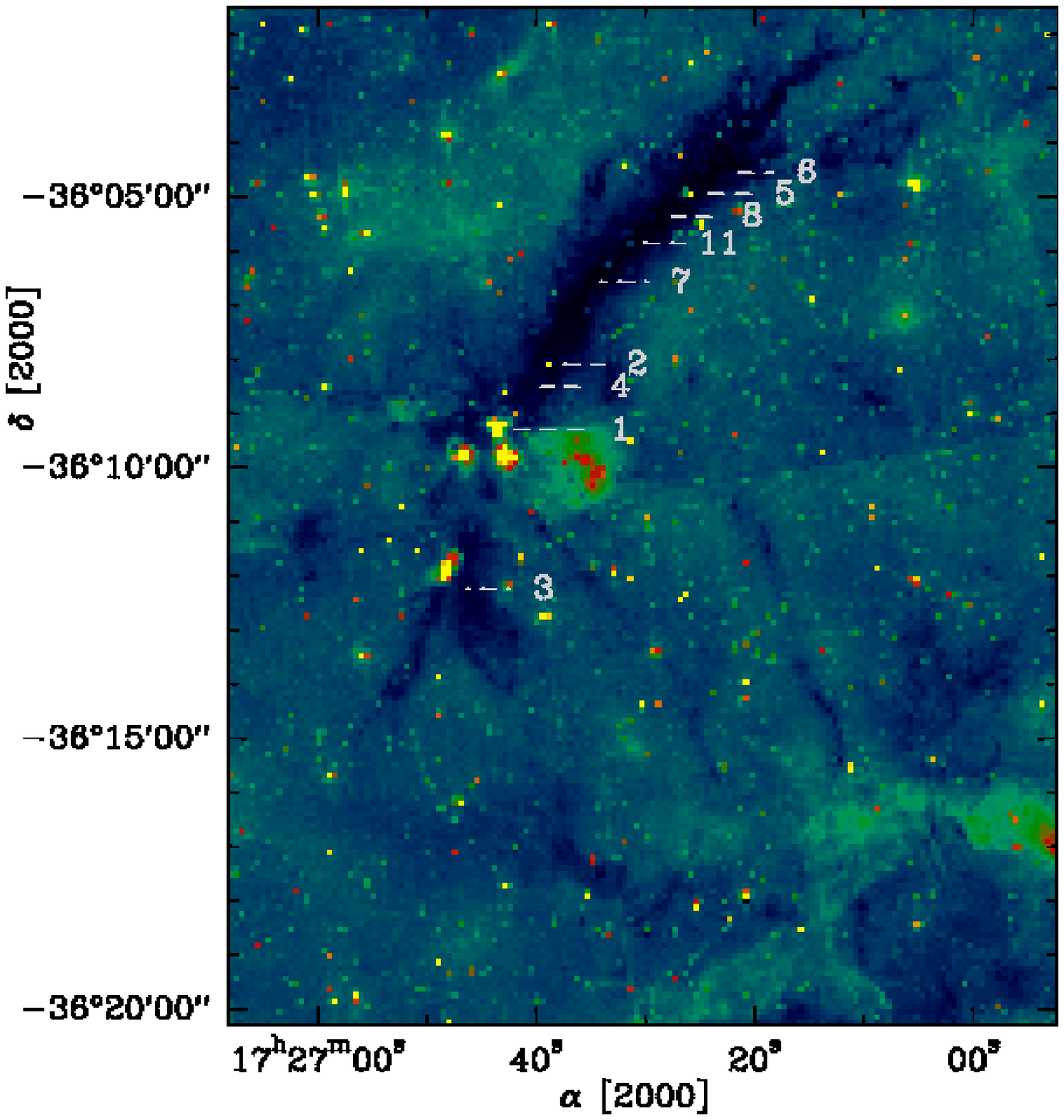}\label{irdc8}}
\subfigure[][]{
\includegraphics[angle=-90, width=9cm]{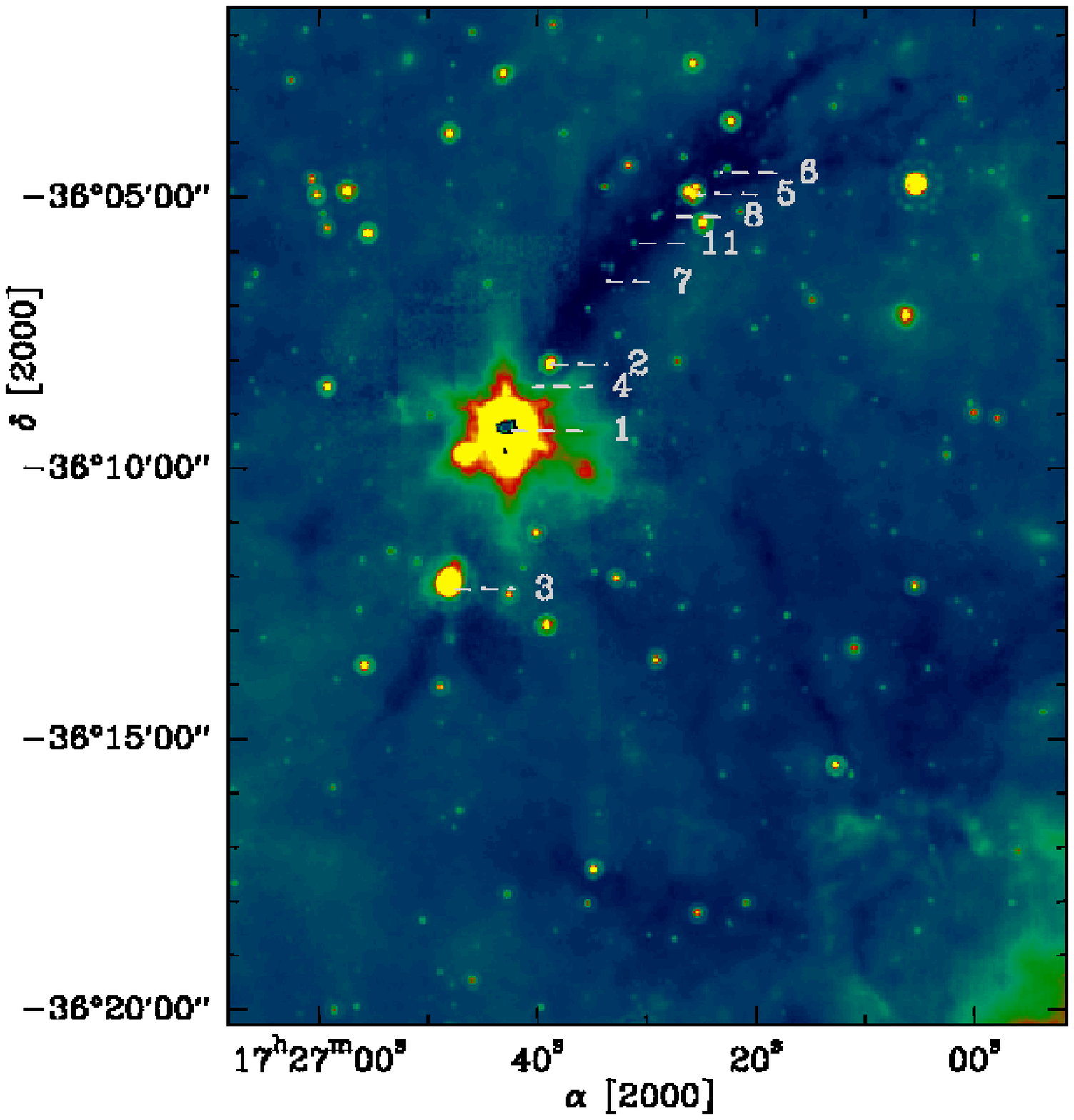}\label{irdc24}}

\caption{Image of the 8$\mu$m ({\it left}) and 24$\mu$m ({\it right}) emission from Spitzer of the molecular
cloud $\rm{G}351.77-0.51$.  The numbers label the 870$\mu$m dust continuum clumps found by the CLUMPFIND
procedure and discussed in the text (see
Table~\ref{clfn}).}\label{irdc}
\end{figure*}

The IRDC under investigation in this paper is at Galactic coordinates
$(l,b)$=$(351^\circ.77,-00^\circ.51)$, and was identified by
\citet{2006ApJ...639..227S} in the 8$\mu$m data from the MSX satellite as  $\rm{G}351.77-0.51$ (hereafter G351.77).
G351.77  appears as a filament of dust, which is seen
in absorption against the mid-IR
background of our Galaxy (see Fig.~\ref{irdc}) up to 24$\mu$m
\citep{2009PASP..121...76C} and in emission in the millimetre range
\citep{2004A&A...426...97F}. Following \citet{2007A&A...476.1243M}, we used the 8$\mu$m MSX image
as a starting point to infer   the stellar activity of the filament.
G351.77 is remarkably quiet at this wavelength: emission is detected towards five positions in the filament, 
three  of them being close to the bright IR source IRAS\,17233$-$3606.
Compact emission is detected in  more
sensitive Spitzer images  towards other positions in the filament (see Sect.~\ref{ir}).
IRAS\,17233$-$3606 lies near one end of the filament.
This source shows several signs of active massive
star formation: different authors
\citep{1980IAUC.3509....2C,1982ApJ...259..657F,1991ApJ...380L..75M}
detected very intense H$_2$O, OH, and CH$_3$OH masers, while
\citet{2008A&A...485..167L,2009A&A...507.1443L} report the discovery of
bipolar outflows originating in the vicinity of IRAS\,17233$-$3606 and a
molecular spectrum typical of hot molecular cores near massive young
stellar objects.     
If  IRAS\,17233$-$3606 is part of the molecular environment of  G351.77,
the distance of 
G351.77 would be relatively close, since studies of IRAS\,17233$-$3606
locate it  at $D\le 1$~kpc \citep[see][and reference therein]{2011arXiv1104.0857L}. 
 The near distance would imply that high linear resolution observations
could be performed even with single-dish radio telescopes whose typical angular resolutions are not
better than $10\arcsec$, corresponding  to 0.05~pc at 1~kpc. In addition, given its low declination, G351.77 would
be an ideal candidate for high angular resolution observations with ALMA to study the earliest stages of formation of massive stars
and stellar cluster, and to investigate the
content of low- and intermediate-mass star formation in IRDCs.

\section{Observations}\label{par-obs}
\subsection{APEX\footnote{Based on
observations collected at the European Organisation for Astronomical
Research in the Southern Hemisphere, Chile, under programme ID
081.F-9805(A). APEX is a collaboration between the
Max-Planck-Institut f\"ur Radioastronomie, the European Southern
Observatory, and the Onsala Space Observatory.}   observations}

The Large Apex BOlomoter CAmera \citep[LABOCA,][]{2009A&A...497..945S} was
used to observe the continuum emission at 870$\mu$m towards the dust
filament G351.77. The {\it FWHM} beam size of APEX at 870$\mu$m is
$\sim 19\farcs2$.  The data  are
part of the ATLASGAL survey of the Galactic plane \citep{2009A&A...504..415S}.
Observations were done with fast-scanning, linear, on-the-fly maps in total-power mode (i.e. without chopping secondary). 
Details on the data reduction are given in \citet{2009A&A...504..415S}. In particular,
the subtraction of correlated skynoise results in a filtering of
{\em{uniform}} extended emission on scales larger than $\sim$2.5$'$;
compact or filamentary sources should not be strongly affected, though. For a detailed discussion of filtering
of extended structures, we refer to \citet{2011A&A...527A.145B}.
The calibration of the LABOCA data relies on opacity correction based on skydips and on observations of planets as primary calibrators and of well-known, bright, compact sources as secondary calibrators. As a result, the photometric uncertainty is of the order of 10\%.

\begin{table}
\centering
\caption{Summary of the parameters of the APEX and MOPRA line observations.}\label{obs}
\begin{tabular}{crccl}
\hline
\hline
\multicolumn{1}{c}{Setup} &\multicolumn{1}{c}{Tuning Frequency}&\multicolumn{1}{c}{Telescope}&
\multicolumn{1}{c}{Beam}&\multicolumn{1}{c}{$\Delta \rm{v}$}\\
\multicolumn{1}{c}{} &\multicolumn{1}{c}{(MHz)}&&\multicolumn{1}{c}{($\arcsec$)}&\multicolumn{1}{c}{(km~s$^{-1}$)}\\
\hline
N$_2$H$^+$ (1--0)&93173.5&MOPRA&$35\farcs0$&0.1\\
$^{13}$CO-C$^{18}$O (2--1)&219980.0&APEX&$28\farcs4$&0.17\\
C$^{17}$O (2--1)&224700.0&APEX&$27\farcs8$&0.16\\
\hline

\end{tabular}

\end{table}
Observations in $^{13}$CO, C$^{18}$O and C$^{17}$O $(2-1)$ towards
selected positions in G351.77 were performed during the science
verification programme (project number 081.F-9805(A)) of the APEX-1
receiver on APEX in 2008 June (19 and 25) and August (21, 30 and 31).
The observations in June were performed under mediocre weather
conditions (with 2--3 mm precipitable water vapour), yielding system
temperatures of 240-400~K. The observations in August were, on the
other hand, performed under excellent atmospheric transmissions (down to
0.4 mm precipitable water vapour), yielding system temperatures of
160-210~K. The observed transitions and basic observational parameters
are summarised in Table~\ref{obs}.

For the observations, we used the APEX facility FFT spectrometer
\citep{2006A&A...454L..29K}.
The $^{13}$CO and C$^{18}$O $(2-1)$ transitions were covered in one
frequency setup, with the APEX-1 receiver tuned to 219.98~GHz in the lower
side band. A second frequency setup was used to cover the C$^{17}$O $(2-1)$ line,
with the APEX-1 receiver tuned to 224.7~GHz in the lower side band.
For both setups, we
performed long integration observations on selected positions
(Table~\ref{pos}). For the $^{13}$CO--C$^{18}$O setup, we observed
an additional strip of five points   (($20\arcsec$,$-20\arcsec$), ($10\arcsec$,$-10\arcsec$), ($0\arcsec$,$0\arcsec$), 
($-10\arcsec$,$10\arcsec$), and ($-20\arcsec$,$20\arcsec$)), centred on the position G351.62 of
Table~\ref{pos}.
For both frequency setups, data were taken in the position-switching mode, with a reference
position  $(600\arcsec, 0\arcsec)$  in horizontal coordinates from the targets.

We used a main-beam efficiency of
0.75\footnote{http://www.apex-telescope.org/telescope/efficiency/index.php}
to convert antenna temperatures into main-beam temperatures. A
detailed description of APEX and of its performance is given by
\citet{2006A&A...454L..13G}.

\begin{table}
\centering
\caption{Positions observed with APEX}\label{pos}
\begin{tabular}{lccc}
\hline
\hline
\multicolumn{1}{c}{Position} &\multicolumn{1}{c}{R.A. [J2000]}&\multicolumn{1}{c}{Dec. [J2000]}&\multicolumn{1}{c}{Setup}\\
\multicolumn{1}{c}{} &\multicolumn{1}{c}{}&\multicolumn{1}{c}{}&\multicolumn{1}{c}{}\\
\hline
Clump-1$^{\mathrm{a}}$ & 17:26:42.30&-36:09:18.23&$^{13}$CO-C$^{18}$O, C$^{17}$O\\
Clump-2  & 17:26:38.79  &   -36:08:05.53&$^{13}$CO-C$^{18}$O, C$^{17}$O\\
Clump-3  & 17:26:47.33  &   -36:12:14.37&$^{13}$CO-C$^{18}$O, C$^{17}$O\\
Clump-4  & 17:26:40.30  &   -36:08:29.63&$^{13}$CO-C$^{18}$O\\
Clump-5  & 17:26:25.25  &   -36:04:57.03&$^{13}$CO-C$^{18}$O, C$^{17}$O\\
Clump-6  & 17:26:23.25  &   -36:04:32.74&$^{13}$CO-C$^{18}$O\\
Clump-7  & 17:26:34.78  &   -36:06:34.22&$^{13}$CO-C$^{18}$O, C$^{17}$O\\
Clump-8  & 17:26:28.26  &   -36:05:21.34&$^{13}$CO-C$^{18}$O\\
Clump-9$^{\mathrm{b}}$  & 17:25:31.96  &   -36:22:15.70&$^{13}$CO-C$^{18}$O, C$^{17}$O\\
Clump-10$^{\mathrm{c}}$ & 17:26:06.66  &   -36:22:09.63&$^{13}$CO-C$^{18}$O, C$^{17}$O\\
Clump-12$^{\mathrm{b}}$ & 17:25:31.97  &   -36:21:21.05&$^{13}$CO-C$^{18}$O\\
G351.62  & 17:26:34.80&-36:17:41.00&$^{13}$CO-C$^{18}$O \\

\hline

\end{tabular}
\begin{list}{}{}
\item[$^{\mathrm{a}}$] IRAS\,17233--3606
\item[$^{\mathrm{b}}$] Clump-9 and 12 are offset from IRAS\,17221--3619
\item[$^{\mathrm{c}}$] IRAS\,17227--3619
\end{list}
\end{table}

\subsection{MOPRA\footnote{The MOPRA telescope
is part of the Australia Telescope which is funded by the Commonwealth of
Australia for operation as a National Facility managed by CSIRO. The University of
New South Wales Digital Filter Bank used for the observations with the MOPRA  
Telescope was provided with support from the Australian Research Council} observations}

The observations were carried out with the ATNF Mopra telescope on 2007
June 30  under clear sky conditions. Two adjacent fields were
observed in $220\times220$ arcsec$^2$ on-the-fly maps for 30 minutes each.  The
HEMT receiver was tuned to 92.2 GHz and the MOPS spectrometer was used
in narrow band mode to cover the 8~GHz
bandwidth of the receiver with about 0.1~km/s velocity resolution. The
beam size at these frequencies is about 35 arcsec. The system
temperature was 175~K.  Pointing was checked and corrected with line
pointings on the SiO maser source AH Sco.

Initial processing of the data was done using the ATNF tools
{\it livedata} and {\it gridzilla} and included the subtraction of the
off signal, the calibration to T$_{a}^*$ units, the baseline subtraction and the
gridding.   To convert antenna to main beam brightness temperature, we corrected the data for an antenna efficiency of
 0.5\footnote{http://www.narrabri.atnf.csiro.au/mopra/obsinfo.html}.
Details on antenna efficiency in the 3mm band is given by \citet{2005PASA...22...62L}.
The data was then exported to MIRIAD and GILDAS for further
processing.

\section{IR emission}\label{ir}

Figures~\ref{irdc8} and \ref{irdc24} show G351.77 at 8 and 24$\mu$m,
respectively, from the GLIMSPE and MIPSGAL surveys of the Galactic
plane \citep{2003PASP..115..953B,2009PASP..121...76C}.  The filament
is dark at both wavelengths except for bright emission associated with
IRAS\,17233--3606 (dust condensation 1, hereafter clump 1) and with the dust condensation 3 (clump 3) to the south of
IRAS\,17233--3606 (see Table~\ref{clfn}).  Additionally, weak emission,
which is still compact in most of the cases, is detected at several positions
along the filament at both wavelengths. Compact emission at 24$\mu$m is an indicator of active star formation,
since this emission traces
warm dust  heated as material accretes from a core onto a central protostar.
To search for evidence of outflow activity from young stellar objects,
we also checked the 4.5$\mu$m map of the region made with the Spitzer InfraRed Array Camera (IRAC).
Extended green emission (4.5$\mu$m) is clearly associated only with the outflows originating in IRAS\,17233--3606
\citep{2009A&A...507.1443L}. A jet-like feature extends from IRAS\,17233--3606 into the dark filament. More
compact emission at 4.5$\mu$m is also detected  towards three additional  positions in G351.77 (clumps 2, 3, and 5).

A larger map of the region in the IRAC 8$\mu$m image reveals a
complex picture: extended emission along a
semicircular structure is detected opposite to G351.77. Moreover,
the map is
dominated by the infrared source
IRAS\,17221--3619, which is associated with bubble CS84 of the catalogue of
\citet{2007ApJ...670..428C}, and  belongs to a second circular structure
seen in emission in the mid-IR (Fig.~\ref{ring}).  
Several dark patches are
seen at 8$\mu$m and 24$\mu$m, two of them being  
the IRDCs
$\rm{G}351.62-0.58$ and  $\rm{G}351.77-0.60$   (see Fig.~\ref{ring} and Table~\ref{pos})
identified in the catalogue of \citet{2006ApJ...639..227S}.

\section{Continuum emission at 870$\mu$m}\label{870micron}

\begin{figure}
\centering
\includegraphics[angle=-90, width=9cm]{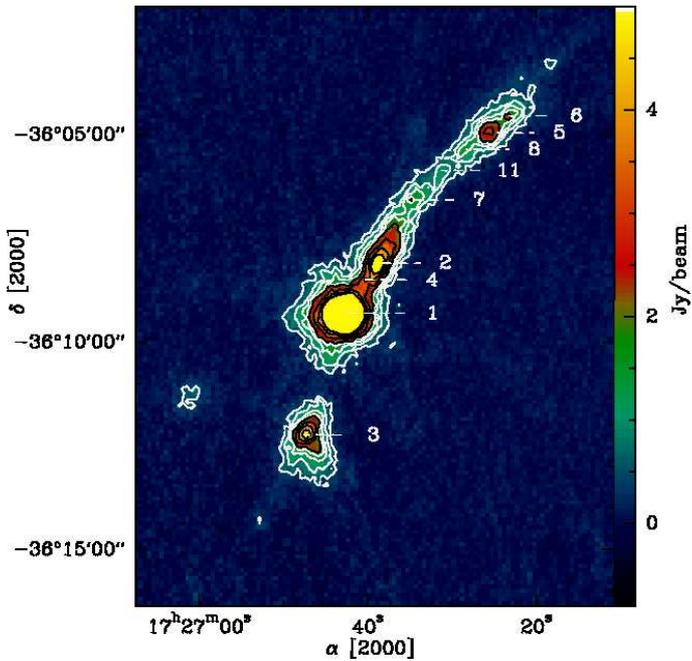}
\caption{LABOCA map of the 870$\mu$m dust continuum emission from the IRDC G351.77--0.51.
The white contours are from 5$\sigma$ (0.425~Jy/beam) to 2~Jy/beam in steps of 5$\sigma$; the black contours are
from 2 ~Jy/beam to 5~Jy/beam in steps of 10$\sigma$.
The numbers label the clumps found by the CLUMPFIND
procedure and discussed in the text (see
Table~\ref{clfn}).}\label{irdc850}
\end{figure}

The continuum emission at 870$\mu$m probes cold dust. Therefore,  it
can be  used to locate dense compact clumps, potential sites
of star and cluster formation. 
Given the linear resolution of our LABOCA map
($19\farcs2\sim 0.1$~pc at 1~kpc), our data are sensitive to dust
clumps possibly associated with cluster formation \citep{2000prpl.conf...97W,2000ApJ...540L..53T,2002A&A...392.1053O}.

\begin{figure*}
\centering
\includegraphics[angle=-90, width=18cm]{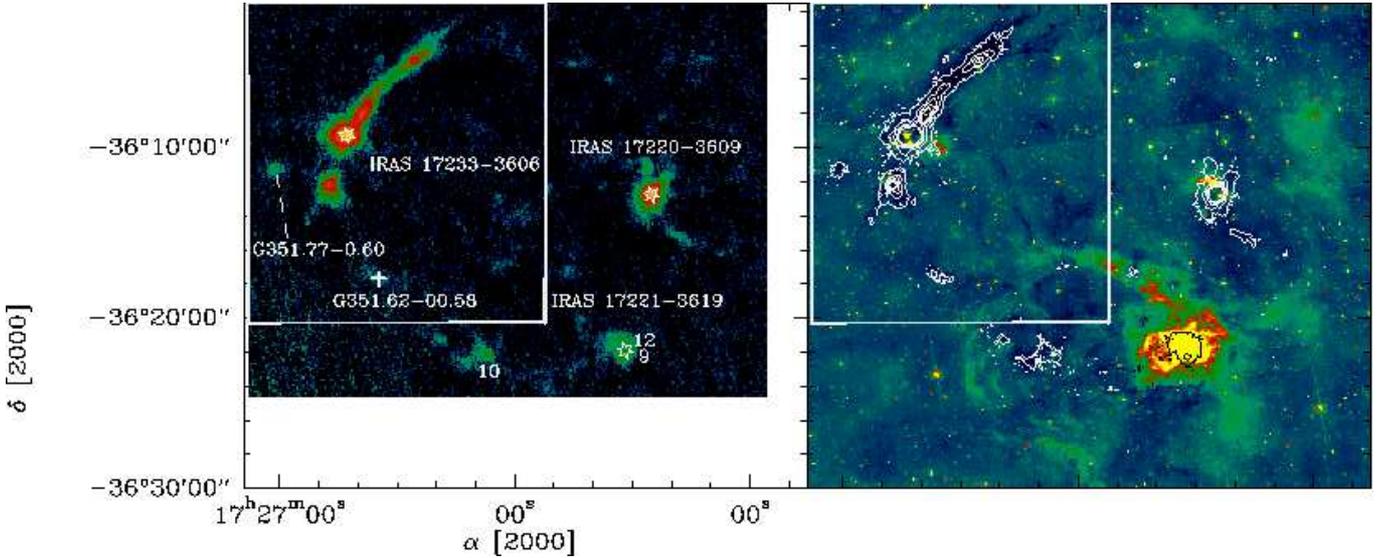}
\caption{{\it Left:} Continuum emission at 870$\mu$m of the environment around  $\rm{G}351.77-0.51$.
The numbers label the clumps found by the CLUMPFIND algorithm outside the region plotted in Fig.~\ref{irdc}.
The white cross marks the position G351.62, the stars  the IRAS sources found in the region.
 {\it Right:}
Image of the emission at 8$\mu$m of the molecular environment around  $\rm{G}351.77-0.51$. The white
(and black for IRAS\,17221--3619) contours show the emission at 870$\mu$m from 0.255 (3~$\sigma$) to 4~Jy/beam in step of 10~$\sigma$.
In both panels, the white box outlines the region plotted in Fig.~\ref{irdc}.
}\label{ring}
\end{figure*}

Our LABOCA map of $\rm{G}351.77$ reveals extended continuum dust
emission that matches the morphology of the mid-IR extinction
extremely well (Figs.~\ref{irdc850} and \ref{ring}). The IRDC $\rm{G}351.77-0.60$ 
coincides with the $6\sigma$ LABOCA emission
on the  east of
IRAS\,17233--3606 (Fig.~\ref{ring}).
On larger scales,
the  emission  extends over a  curved structure (Fig.~\ref{ring}),
which includes the IRDC $\rm{G}351.62-0.58$ (see previous section) associated with our position G351.62,
and the sources  IRAS\,17227--3619 and IRAS\,17221--3619 (see Table~\ref{pos}).
A second, intense peak of continuum emission is found opposite to IRAS\,17233--3606, on its west.
This emission is, however, associated with the
IR source IRAS\,17220--3609, whose molecular lines have a velocity of $-98.5$~km~s$^{-1}$ \citep{1996A&AS..115...81B},
 which implies a kinematic distance of
10~kpc \citep{2004A&A...426...97F}.
Since IRAS\,17233--3606 has a velocity of
$-3.4$~km~s$^{-1}$, the two sources
are unlikely to be physically related. We conclude that IRAS\,17220--3609 is projected close to the
line of sight of IRAS\,17233--3606,
but it is much further away from the Sun, so we did not include
IRAS\,17220--3609 in our following analysis.

No continuum emission at 870$\mu$m is associated with the
semicircular structure opposite to G351.77 seen at 8$\mu$m (Fig.~\ref{ring}),
hence probably due to low density
gas not sampled by our observations. Since our molecular line
follow-up observations were only performed toward peaks of emission at
870$\mu$m, we do not have any information on the velocity field of
this structure and we cannot determine whether it is physically associated
with G351.77 or not.

\subsection{Identification of the clumps}\label{870}
\begin{table}
\centering
\caption{Results of the CLUMPFIND calculations}\label{clfn}
\begin{tabular}{crrrrrc}
\hline
\hline
\multicolumn{1}{c}{Clump} &\multicolumn{1}{c}{$S_\nu$}&\multicolumn{1}{c}{$F_\nu$}&
\multicolumn{1}{c}{$\theta_x^{\mathrm{a}}$}&\multicolumn{1}{c}{$\theta_y^{\mathrm{a}}$}&\multicolumn{1}{c}{$\theta^{\mathrm{b}}$}&
\multicolumn{1}{c}{IR$^{\mathrm{c}}$}\\
\multicolumn{1}{c}{}&\multicolumn{1}{c}{[Jy]}  &\multicolumn{1}{c}{[Jy/beam]}  &\multicolumn{1}{c}{[$\arcsec$]}&\multicolumn{1}{c}{[$\arcsec$]}&\multicolumn{1}{c}{[$\arcsec$]}\\
\hline
\multicolumn{7}{c}{$\rm{G}351.77-0.51$}  \\
1  & 159.4 & 54.2 &43.6&38.5&41.1&$8, 24\mu$m\\
2  &  29.7 & 6.2  &31.6&46.7&39.3&$8, 24\mu$m\\
3  &  24.9 & 5.2  &31.5&49.6&40.7&$8, 24\mu$m$^{\mathrm{d}}$ \\
4  &  16.1 & 4.0  &46.9&25.2&36.5&$^{\mathrm{e}}$ \\
5  &   9.3 & 3.1  &18.8&20.9&19.8&$8, 24\mu$m  \\
6  &   6.1 & 2.3  &21.5&19.3&20.4&$24\mu$m  \\
7  &   8.7 & 2.1  &33.0&41.5&37.3&\\
8  &   4.6 & 1.5  &17.5&20.5&19.1&$24\mu$m \\
11 &   3.2 &1.1   &20.6&22.7&21.7&$8, 24\mu$m\\                                                 
\multicolumn{7}{c}{other positions} \\               
9  &  4.8 &1.5 &28.9&21.1&25.1&$8, 24\mu$m\\  
10 &  2.7 &1.2 &23.1&12.4&18.2&$8, 24\mu$m \\  
12 &  5.0 &1.0 &46.9&21.5&34.8&$8, 24\mu$m\\

\hline

\end{tabular}
\begin{list}{}{}
\item[$^{\mathrm{a}}$] deconvolved ${{\it FWHM}_x}$ and ${{\it FWHM}_y}$
\item[$^{\mathrm{b}}$] deconvolved geometric mean of ${{\it FWHM}_x}$ and ${{\it FWHM}_y}$
\item[$^{\mathrm{c}}$] association with IR emission
\item[$^{\mathrm{d}}$] the IR source is offset from the mm source
\item[$^{\mathrm{e}}$] position contaminated by the emission from IRAS\,17233--3606
\end{list}
\end{table}

To identify  clumps in the millimetre continuum emission and define their
properties, we  used a
two-dimensional variation of the clump-finding algorithm CLUMPFIND
developed by \citet{1994ApJ...428..693W}.  
The algorithm works by effectively contouring the data at a
multiple of the rms noise of the map, then searching for peaks of
emission to locate the clumps, and finally following the clump profile
down to lower intensities. The advantage of this method is that it does
not a priori assume any
shape for the clumps. However,
since the contouring levels are chosen by hand,
the
algorithm does not take any emission into account below the user-selected threshold.
Recently, \citet{2009ApJ...699L.134P}  have outlined some weaknesses of the method, which may lead to
erroneous mass functions for crowded regions. Nevertheless, the CLUMPFIND algorithm may still be useful
for studying the structure of one cloud as in our case.
We set the threshold level
to 5$\sigma $. 

The procedure calculates the peak position, the full width at half
maximum  not corrected for beam size for the x-axis, ${{\it FWHM}_x}$, and
for the y-axis, ${{\it FWHM}_y}$, and the total flux density integrated within
the clump boundary  within the threshold level.

We identified twelve clumps in the 870$\mu$m continuum emission over the region shown in Fig.~\ref{ring}.
The centroid coordinates delivered by CLUMPFIND are given in Table~\ref{pos}. Other results of the CLUMPFIND analysis are given in Table~\ref{clfn}
where we also  indicate whether clumps are
associated with continuum emission at 8 and 24$\mu$m, either point-like or diffuse.
We interpret the detection of a compact 24$\mu$m source in all clumps of G351.77 except clumps 4
and 7 as a confirmation of
the validity of the method used for identifying clumps in the region.
While clump 7 is clearly detected as a secondary peak of emission at 870$\mu$m, clump 4 does not look like
 a clumpy structure at visual inspection, and could be an example of 'pathological' features {\it created} by
CLUMPFIND in case of significant emission found between prominent clumps \citep[see][]{2009Natur.457...63G}.
Additionally, we note that
clump~1 is associated with the
active massive star-forming region IRAS\,17233--3606 discussed above, and that clumps~9 and 12 are associated
with the infrared source IRAS\,17221--3619, which hosts an H{\sc ii} region
\citep[e.g.,][]{2003A&A...407..957M} surrounding an IR star cluster candidate \citep{2006A&A...455..923B} associated with bubble CS84.

\begin{table*}
\begin{center}
\caption{Physical parameters of the clumps derived from the LABOCA emission.}\label{clfn-ph}
\begin{tabular}{cccccccccc}
\hline
\hline
\multicolumn{1}{c}{Clump} &\multicolumn{1}{c}{$M_{\rm{10K}}$}&\multicolumn{1}{c}{$M_{\rm{25K}}$}&\multicolumn{1}{c}{$M_{\rm{35K}}$}
&\multicolumn{1}{c}{$N_{{\rm{H}_2},{\rm 10K}}$}&\multicolumn{1}{c}{$N_{{\rm{H}_2},{\rm{25K}}}$}&\multicolumn{1}{c}{$N_{{\rm{H}_2},{\rm 35K}}$}
&\multicolumn{1}{c}{$n_{{\rm{H}_2},{\rm 10K}}$}&\multicolumn{1}{c}{$n_{{\rm{H}_2},{\rm{25K}}}$}&\multicolumn{1}{c}{$n_{{\rm{H}_2},{\rm 35K}}$}\\
&\multicolumn{1}{c}{[$M_\odot$]}&\multicolumn{1}{c}{[$M_\odot$]}&\multicolumn{1}{c}{[$M_\odot$]}&\multicolumn{1}{c}{[$10^{22}$~cm$^{-2}$]}&\multicolumn{1}{c}{[$10^{22}$~cm$^{-2}$]}&\multicolumn{1}{c}{[$10^{22}$~cm$^{-2}$]}&\multicolumn{1}{c}{[$10^5$~cm$^{-3}$]}&\multicolumn{1}{c}{[$10^5$~cm$^{-3}$]}&\multicolumn{1}{c}{[$10^5$~cm$^{-3}$]}\\
\hline
\multicolumn{10}{c}{$\rm{G}351.77-0.51$}  \\
1   &  2964 & 664 & 428 & 460 & 103 & 66 & 108 &  24 &  16\\
2   &   552 & 124 &  80 &  52 &  12 &  8 &  23 &   5 &   3\\ 
3   &   463 & 104 &  67 &  44 &  10 &  6 &  17 &   4 &   3\\
4   &   300 &  67 &  43 &  34 &   8 &  5 &  16 &   3 &   2\\
5   &   173 &  39 &  25 &  27 &   6 &  4 &  56 &   13 &   8\\
6   &   113 &  25 &  16 &  19 &   4 &  3 &  34 &   7 &   5\\
7   &   161 &  36 &  23 &  18 &   4 &  3 &   8 &   2 &   1\\
8   &    85 &  19 &  12 &  13 &   3 &  2 &  31 &   7 &   4\\
11  &    60 &  13 &   9 &  10 &   2 &  1 &  15 &   3 &   2\\

\multicolumn{10}{c}{other positions} \\               
9   &    89 &  20 &  13 &  13 &   3 &  2 &  14 &   3 &   2\\
10  &    51 &  11 &   7 &  10 &   2 &  1 &  21 &   5 &   3\\
12  &    94 &  21 &  14 &   9 &   2 &  1 &   6 &   1 &   1\\
 
\hline
\end{tabular}
\end{center}
\end{table*}

\subsection{Mass estimate}
The masses of the clumps given in Table~\ref{clfn-ph} are derived by assuming that the continuum
emission at 870$\mu$m is optically thin:

\begin{equation}
M_{tot}=\frac{d^2 F_\nu}{\kappa_\nu B_\nu(T_d)}
\end{equation}
where $F_\nu$ is the integrated flux, $d$ the distance of the source,
$\kappa$ the dust opacity, and $B(T_d)$ is the Planck function for a
black body of dust temperature $T_d$, all for a frequency of $\sim 340$
GHz.  For the dust opacity,
we adopted a value of 0.0182 cm$^{2}$~g$^{-1}$
\citep[from Table 1, Col. 5 of][]{1994A&A...291..943O}. This agrees with
recent measurements of dust opacities at 850 and 450$\mu$m from \citet{2011ApJ...728..143S} in the low-mass Class 0 core B335. 

From the line observations performed towards the dust filament of G351.77, we confirm
that IRAS\,17233--3606 belongs to the same molecular environment of the infrared dark cloud G351.77 (see Sect.~\ref{co}).
 We assume a distance of 1~kpc in the following calculations. 
Our mass estimates would be a factor of 2 smaller for a distance of 700~pc (see Sect.~\ref{intro}), corresponding to the smallest distance 
of IRAS\,17233--3606 reported in the literature \citep{2006A&A...460..721M}.
 We also computed
 beam-averaged H$_2$ volume densities assuming that
the clumps have spherical symmetry and a mean molecular weight per hydrogen molecule, $\mu_{\rm H_2}$, of 2.8 
 (see discussion in \citealt{2008A&A...487..993K}). 
From previous line observations, we know that clump 1 (IRAS
17233--3606) harbours a hot core \citep{2008A&A...485..167L}, while our
current data towards the other clumps do not reveal any rich
molecular spectrum indicative of hot core activity, i.e., an embedded central heating source.
 Ammonia (1,1) and (2,2) observations with the Parkes telescope
(Wienen et al. in prep.) of three positions along the filament (clump 1, 2, and 3) infer temperatures of  $<$20~K on
an angular size of $\sim1\arcmin$.
Except for clump 1, which hosts the hot core associated with IRAS\,17233--3606,
 we believe that temperatures of 10--25~K are appropriate for the
dust in the clumps
in G351.77 where only point-like emission is detected at 24$\mu$m.
 We computed masses, column densities, and densities of all clumps also for $T_d=35$~K for consistency with Sect.~\ref{colden}.
 Results can be easily scaled
to higher temperatures: compared to the values given for 
$T_d=35$~K, estimates are a factor $\sim1.5$ smaller for $T_d=50$~K, likely to be an upper limit to the
temperature of the dust for IRAS\,17233--3606 at the resolution of the data. 
For $T=25$~K,
the lowest mass derived from our analysis for the clumps in G3512.77 is 13~$M_\odot$ (clump~11), while all other clumps
have masses of several tens of solar masses. As a result, G351.77 has the potential of forming intermediate to high-mass stars or
clusters of low-mass stars
in several positions along the filament.

\section{Emission from CO isotopologues}\label{co}

\subsection{Velocity field}
\begin{figure*}
\centering
\includegraphics[angle=-90, width=15cm]{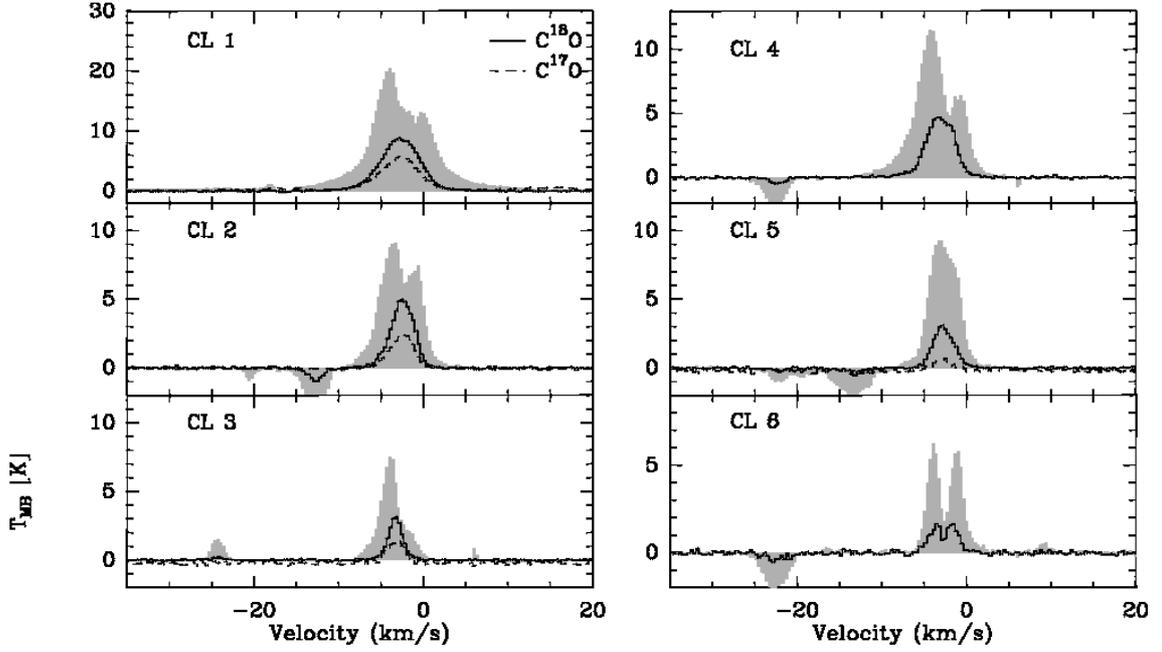}
\caption{In grey, spectra of $^{13}$CO (2--1)  of clumps 1, 2, 3, 4, 5 and 6. The solid and dashed spectra
are C$^{18}$O (2--1), and C$^{17}$O (2--1) respectively.}\label{1_6}
\end{figure*}

\begin{figure*}
\centering
\includegraphics[angle=-90, width=15cm]{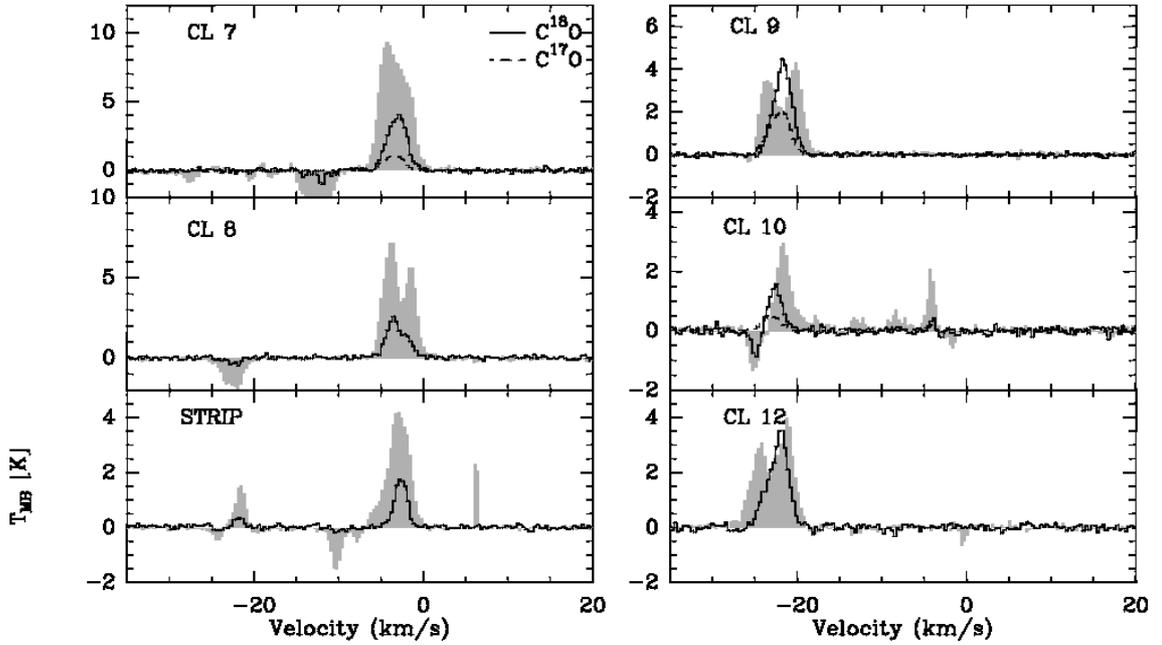}
\caption{In grey, spectra of $^{13}$CO (2--1) of clumps 7, 8, 9, 10, 12 and position G351.62. The solid and dashed  spectra
are C$^{18}$O (2--1), and C$^{17}$O (2--1) respectively.}\label{7_st}
\end{figure*}

To derive the velocity of each clump, we used the optically thin
C$^{17}$O $(2-1)$ line (see Sect.~\ref{colden}). The C$^{18}$O $(2-1)$ transition was used for the
positions that are not observed in C$^{17}$O (see Table~\ref{pos}).  Both transitions
were detected towards all  positions towards which they were observed.
To derive the parameters of the C$^{17}$O line, we used the HFS method in CLASS. A Gaussian fit
was applied to the  C$^{18}$O and $^{13}$CO $(2-1)$ lines,
although  their spectra show more complex
velocity structures. The $^{13}$CO line often shows evidence of non-Gaussian wings and
multiple peaks. These are probably due to self absorption, as the optical thinner C$^{18}$O and
C$^{17}$O transitions peak at  the velocity of the dip in the $^{13}$CO spectrum.
Negative features at other velocities are most likely due to emission in the reference positions used
during the observations. However, in the clumps that were observed in all three isotopologues,
no absorption feature is detected in the C$^{17}$O spectrum. Therefore we believe that our results (based on
the analysis of the C$^{17}$O spectra, see Sect.~\ref{co_n2h+}) are not affected by
this problem.
The spectra of all clumps are shown in Figs.~\ref{1_6} and \ref{7_st}, while the results of
the fit are listed in Table~\ref{others}.

From the analysis of the velocity field, we found two different molecular
components along the ring of dust detected at $870~\mu$m, and all clumps along the
infrared dark cloud G351.77 are at velocities $-3$~km~s$^{-1}$,  including clump~3, immediately
south of IRAS\,17233--3606 (clump~1). We confirm the $5\sigma$ dust continuum detection  
at the position G351.62 $(\alpha_{2000}=17^h 26^m 34^s.80,\delta_{2000}=-36^\circ 17'41\farcs00)$ to be due to
dense material, since both  $^{13}$CO $(2-1)$ and  C$^{18}$O $(2-1)$ are detected at that position, with a velocity
similar to that of G351.77. Clumps 9 and 12 (see Fig.~\ref{ring}) have velocities around $-22$~km~s$^{-1}$.
Moreover, at three positions (clump~3, clump~10,
and G351.62) two velocity components are detected in emission, in
$^{13}$\,CO\,$(2-1)$ at clumps~3 and 10, and in $^{13}$\,CO\,$(2-1$) and C$^{18}$O $(2-1)$
at all five points across the position G351.62.  The two velocity
components (clump~3\footnote{velocities of clumps 3 and 10 are here derived from the $^{13}$CO $(2-1)$ 
line since the second component is not detected in C$^{18}$O}: $-3.2$ and
$-24.5$~km~s$^{-1}$; clump~10: $\sim-4.0$ and
$-21.5$~km~s$^{-1}$; G351.62:  $-2.6$ and $-21.9$~km~s$^{-1}$)
belong to the
same molecular environment of the infrared dark cloud G351.77 and to that of clumps
9 and 12.
On clump~4, the feature at $\rm{v}_{LSR} -3.2$~km~s$^{-1}$ shows non-Gaussian blue-shifted wings; on clump~10, the feature at
$-21.5$~km~s$^{-1}$ also has non Gaussian red-shifted wings.

\begin{table*}
\caption{Line parameters of the  C$^{17}$O and C$^{18}$O$(2-1)$  transitions. The symbol -- means that the line was not observed.}\label{others}
\begin{center}
\begin{tabular}{lcccccc}
\hline
\hline
\multicolumn{1}{c}{Clump} &\multicolumn{1}{c}{$\rm{v}_{\rm LSR}$}&\multicolumn{1}{c}{$\Delta \rm{v}$}&\multicolumn{1}{c}{$\int T_{\rm mb}\delta \rm{v}$$^{\mathrm{a}}$}&\multicolumn{1}{c}{$\rm{v}_{\rm LSR}$}&\multicolumn{1}{c}{$\Delta \rm{v}$}&\multicolumn{1}{c}{$\int T_{\rm mb}\delta \rm{v}$}\\
\multicolumn{1}{c}{}      &\multicolumn{1}{c}{[km~s$^{-1}$]}&\multicolumn{1}{c}{[km~s$^{-1}$]}&\multicolumn{1}{c}{[K~km~s$^{-1}$]}& \multicolumn{1}{c}{[km~s$^{-1}$]}&\multicolumn{1}{c}{[km~s$^{-1}$]}&\multicolumn{1}{c}{[K~km~s$^{-1}$]}\\
\hline
\multicolumn{1}{c}{}&\multicolumn{3}{c}{C$^{17}$O}&\multicolumn{3}{c}{C$^{18}$O}\\

1$^{\mathrm{b}}$    & $-2.51\pm 0.04$&$5.0 \pm 0.1$&$32.1\pm 0.5$      &$-2.65\pm0.02$  &$5.68\pm0.04$&$84.0\pm0.5$\\    
2                   & $-2.21\pm0.02 $&$2.4 \pm 0.1$&$8.4 \pm0.1 $      &$-2.28\pm0.02$  &$3.33\pm0.06$&$21.3\pm0.3$\\    
3                   & $-2.95\pm0.06 $&$1.4 \pm 0.2$&$4.2 \pm0.2$       &$-3.16\pm0.02$  &$2.17\pm0.05$&$10.1\pm0.2$\\      
4                   &  --            & --          &          --       &$-3.16\pm0.02$  &$3.66\pm0.06$&$20.8\pm0.3$\\     
5                   & $-2.6\pm 0.1 $ &$2.2\pm0.3 $ &$2.3 \pm0.2$       &$-2.66\pm0.03$  &$2.82\pm0.07$&$11.7\pm0.3$\\
6$^{\mathrm{c}}$    &  --            & --          &          --       &$-2.5\pm0.1$    &$4.0\pm0.3$  &$6.9\pm0.4$\\
7                   & $-3.35 \pm0.03$&$1.9\pm 0.2 $&$3.52\pm0.07$      &$-3.19\pm0.05$  &$2.6\pm 0.1$ &$12.1\pm 0.4$\\
8$^{\mathrm{d}}$    &  --            & --          &          --       &$-3.39\pm0.05$  &$2.5\pm 0.1$ &$6.8\pm0.3$\\
9                   &$ -21.80\pm0.02$&$1.94\pm0.09$&$6.33\pm0.08$      &$-21.70\pm0.02$ &$2.43\pm0.05$&$14.9\pm0.3$\\
10                  &$ -22.80\pm0.07$&$2.1\pm0.2$  &$1.39\pm0.08$      &$-22.67\pm0.08$ &$1.9\pm 0.2$ &$4.5\pm0.3$\\
12$^{\mathrm{e}}$   &  --                          &         --  &  -- &$-22.27\pm 0.04$&$3.1\pm 0.1$ &$11.1\pm0.3$\\
G351.62$^{\mathrm{f}}$& --                           &         --  &  -- &$-2.58\pm0.03$  &$1.66\pm0.07$&$4.2\pm0.2$\\  
                    & --                           &         --  &  -- &$-22.0\pm0.1$   &$1.3\pm0.2$  &$0.8\pm0.1$\\
\hline
\end{tabular}
\end{center}
Uncertainties are the errors in the Gaussian and/or HFS fit,
and do not include  calibration uncertainties.
\begin{list}{}{}
\item[$^{\mathrm{a}}$] the C$^{17}$O$(2-1)$ integrated intensity has been calculated from a Gaussian fit.
\item[$^{\mathrm{b}}$] the C$^{18}$O$(2-1)$ line has a non Gaussian red- and blue-shifted emission.
\item[$^{\mathrm{c}}$] the C$^{18}$O$(2-1)$ line has a double peak profile. The fit was performed with one component
\item[$^{\mathrm{d}}$] the C$^{18}$O$(2-1)$ line has a non Gaussian red-shifted emission.
\item[$^{\mathrm{e}}$] the C$^{18}$O$(2-1)$ line has a non Gaussian blue-shifted emission.
\item[$^{\mathrm{f}}$] the C$^{18}$O$(2-1)$ line has two velocity components (as well as the $^{13}$CO $(2-1)$ transition);
the one at $\rm{v}_{\rm LSR}=-2.6$ has non Gaussian
blue-shifted  emission.
\end{list}
\end{table*}

\subsection{Column density  estimate}\label{colden}

 From the integrated intensity of the C$^{17}$O $(2-1)$ line, or
alternatively from the C$^{18}$O $(2-1)$ transition for those sources
not observed in C$^{17}$O, we can derive the beam-averaged column density of CO,
assuming a given abundance of CO relative to its rarer isotopologues
\citep[$X_{^{16}{\rm O}/^{18}{\rm O}}\sim 560$, $X_{^{16}{\rm O}/^{17}{\rm O}}\sim 1800$,][]{1994ARA&A..32..191W}, and a given
excitation temperature. Although the HFS method in CLASS infers the
opacities for each components of the hyperfine structure of a
transition, the results are very uncertain for overlapping
components. This is  the case for the C$^{17}$O $(2-1)$ line, where
the largest separation in velocity between the different components is
$\sim 2.4$~km~s$^{-1}$ and lines have typical widths of $\sim
2$~km~s$^{-1}$. 

We therefore computed the C$^{17}$O column density
under the assumption that the $(2-1)$ line is optically thin and that
the gas is in LTE \citep{2002ApJ...565..344C}.  
\citet{2000ApJ...536..393H} found rotational temperatures for C$^{17}$O between 16 and 41~K in a sample
of ultra-compact H{\sc II} regions. Therefore, we computed column densities for $T_{\rm{ex}}=10, 25$ and 35~K. 
 The partition function 
$Q(T_{\rm ex})$
was estimated as $Q(T_{\rm ex})= \alpha T^{\beta}$, where $\alpha$ and
$\beta$ are the best-fit parameters from a fit to the partition
function obtained from CDMS catalogue \citep{2001A&A...370L..49M} at
different excitation temperatures in the range 10--500 K.  The values of
$Q(T_{\rm ex})$ at 10, 25, and 35~K are 3.8, 9.4, and 13.2, respectively for
C$^{17}$O, 3.9, 9.6 and 13.5 for C$^{18}$O.  The results are given in
Tables~\ref{cd} and \ref{cd_18}.  The uncertainties 
are statistical errors based on the fit (see Table~\ref{others}) and do not take the uncertainties 
on the excitation temperature into account. We refer the reader to
Sect.~\ref{co_n2h+} where the effect of an uncertain excitation
temperature on column density is
considered when estimating molecular abundances. 
Moreover, the column densities derived for the two
isotopologues of CO can be lower limits to the real values in case
the lines are optically thick. 

For the C$^{18}$O$(2-1)$ line we can verify
if the assumption of optically thin emission is correct for those clumps that have been
observed in both isotopologues.
The expected ratio $R^{18,17}$
between the integrated intensities of the two transitions
should be similar to the canonical value of 3.65 \citep{1981ApJ...249..518P,1994ARA&A..32..191W}
for the abundance of C$^{18}$O relative to C$^{17}$O. We found
$R^{18,17}>3$ for clumps 7 and 10,
while clumps 1, 2, 3, and 9
have values of $\sim2.5$, which implies that the C$^{18}$O $(2-1)$ transition is partially optically thick.
In addition, clump 5 has  $R^{18,17}\sim5.3$, which could imply that the two isotopologues may have
different excitations or extent of emission. We therefore conclude that the  column densities derived from
the C$^{18}$O transition are likely lower limits to the real values.

\section{Emission from N$_2$H$^+$}

\begin{figure*}
\centering
\includegraphics[width=15cm]{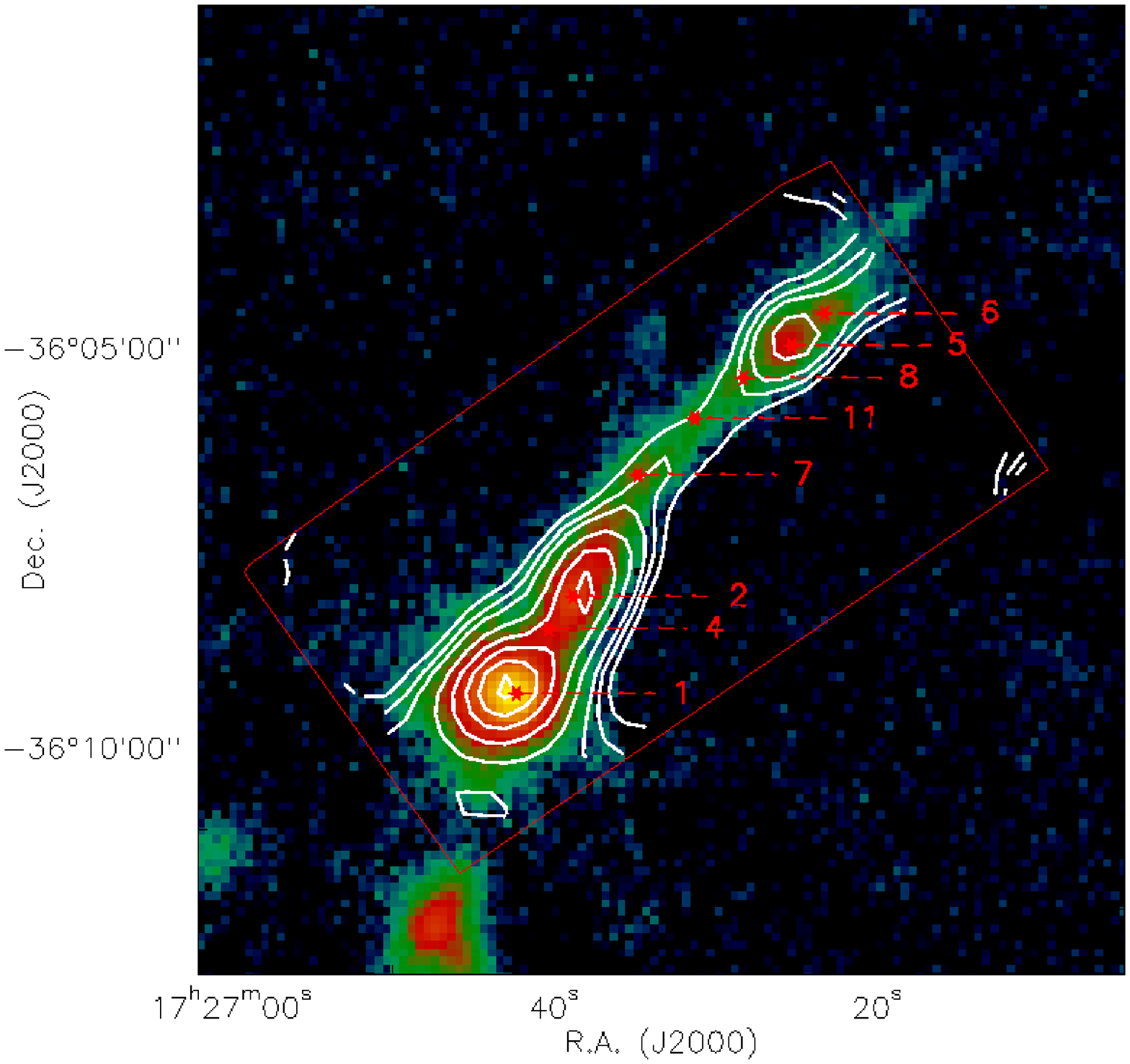}
\caption{MOPRA N$_2$H$^+$ moment zero map in contours overlaid on the LABOCA 870$\mu$m
emission. The red rectangle shows the approximate area mapped with
MOPRA. Clump labelling is the same as adopted through out the text. The
contours have been chosen to start at 3~$\sigma$, 5~$\sigma$ in steps
of 3~$\sigma$ where $\sigma=1.4~{\rm Jy~kms^{-1}}$.\label{fig_n2h+_dust_co}}
\end{figure*}

The lower excitation lines of N$_2$H$^+$ have been shown to be excellent tracers of cold and dense gas
(\citealt{2004A&A...416..191T}, \citealt{2002ApJ...570L.101B}).
With the goal of deriving the properties of the dense gas, we 
mapped the infrared dark patch of G351.77 in N$_2$H$^+$\,(1\,--\,0).
As expected, dust and N$_2$H$^+$ integrated emission correlate very well (see
Fig.~\ref{fig_n2h+_dust_co}). Though the resolution of the MOPRA data
is approximately two times lower than that of our dust continuum data, at least three
bright cores can be identified along the filament. These coincide with the
dust clumps labelled 1, 2, and 5. Clumps 4, 6, 7, and 8 are also
associated with prominent N$_2$H$^+$ emission, although not clearly resolved into cores.
Clump 11 is associated with a ridge of low-level N$_2$H$^+$ emission, consistent with the low dust mass of this clump.

\begin{figure}
\centering
\includegraphics[angle=0,bb=85 2 530 1065,clip, width=8.2cm]{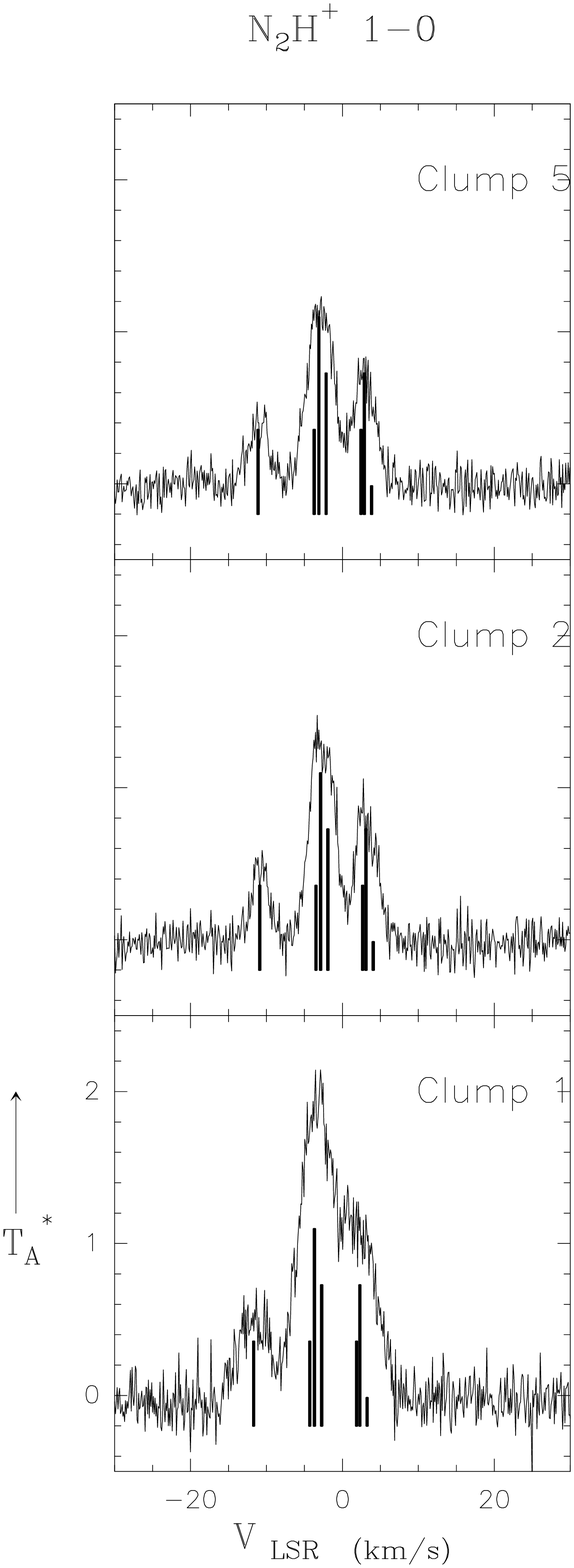}
\caption{Spectra of N$_2$H$^+$ 1--0 of clumps 1, 2, and 5 (clumps
  identified by CLUMPFIND).  The black lines in each spectrum mark the seven
    hyperfine components.}\label{n2hspec}
\end{figure}

The interaction of the molecular electric field gradient and the electric
quadrupole moments of the nitrogen nuclei (I=1) leads to hyperfine structure
and a splitting of the N$_2$H$^+$  $(1-0)$  transition into seven components.
The large line widths often attributed to the high level of turbulence in high-mass star-forming
regions, lead to blending and result in three blended groups (one main
and two satellites). As in the case of C$^{17}$O, the CLASS HFS method has
been used to fit the line profile. Thus, we can accurately determine
the line optical depth, excitation temperature, line width, and LSR
velocity. These parameters were determined for the four clumps
averaged over the MOPRA beam of 35\arcsec.
Subsequently, we  determined the line column density following
\citet{2002ApJ...565..344C}. The spectra of the three clumps are
 shown in Fig.~\ref{n2hspec}, while the line parameters are tabulated
 in Table~\ref{tab_n2h+}. The heavy blending of hyperfine components means 
less stringent constraints on the optical depth and excitation
temperature.  In particular, the optical depth for clump 1 derived from the HFS method in CLASS is unexpectedly low (0.2)
with $>50$\% error. \ The fit to the hyperfine structure is as good
for a range in optical depth, $0.2<\tau_{\rm tot}<1$. Hence, in Table~\ref{tab_n2h+} we give
the range in column densities, abundance and abundance ratio bracketing $\tau_{\rm tot}=0.2$ and $\tau_{\rm tot}=1$. Fits for $\tau_{\rm tot}>1$ were poorer and discarded.\\

\begin{table}
\centering
\caption{Column densities of C$^{17}$O.}\label{cd}
\begin{tabular}{lrrr}
\hline
\hline
\multicolumn{1}{c}{Clump} &\multicolumn{3}{c}{$N_{\rm{C^{17}O}}$}\\
\multicolumn{1}{c}{} &\multicolumn{3}{c}{($10^{15}$~cm$^{-2}$)}\\
&\multicolumn{1}{c}{$T_{\rm{ex}}=10$~K}&\multicolumn{1}{c}{$T_{\rm{ex}}=25$~K}&\multicolumn{1}{c}{$T_{\rm{ex}}=35$~K}\\

1 &$38.8\pm0.6$&$35.6\pm0.6$&$41.3\pm0.6$\\   
2 &$10.2\pm0.1$&$9.3\pm0.1$&$10.8\pm0.1$\\
3 &$5.1\pm0.2$&$4.7\pm0.2$&$5.4\pm0.3$\\  
5 &$2.8\pm0.2$&$2.6\pm0.2$&$3.0\pm0.3$\\
7 &$4.25\pm0.08$&$3.91\pm0.08$&$4.5\pm0.1$\\
9 &$7.7\pm0.1$&$7.03\pm0.09$&$8.1\pm0.1$\\
10&$1.7\pm0.1$&$1.54\pm0.09$&$1.8\pm0.1$\\

\hline
\end{tabular}
\end{table}
\begin{table}
\begin{center}
\caption{Column densities of C$^{18}$O.}\label{cd_18}
\begin{tabular}{lrrr}
\hline
\hline
\multicolumn{1}{c}{Clump} &\multicolumn{3}{c}{$N_{\rm{C^{18}O}}$}\\
\multicolumn{1}{c}{} &\multicolumn{3}{c}{($10^{15}$~cm$^{-2}$)}\\
\multicolumn{1}{c}{}&\multicolumn{1}{c}{$T_{\rm{ex}}=10$~K}&\multicolumn{1}{c}{$T_{\rm{ex}}=25$~K}
&\multicolumn{1}{c}{$T_{\rm{ex}}=35$~K}\\
1&$104.0\pm0.6$&$97.5\pm0.6$&$113.4\pm0.7$\\
2&$26.4\pm0.4$&$24.7\pm0.3$&$29.6\pm0.4$\\
3&$12.5\pm0.2$&$11.7\pm0.2$&$13.6\pm0.3$\\
4&$25.6\pm0.3$&$24.1\pm0.3$&$28.1\pm0.4$\\
5&$14.5\pm0.5$&$13.6\pm0.5$&$15.8\pm0.5$\\
6&$8.5\pm0.5$&$8.0\pm0.5$&$9.3\pm0.5$\\
7&$15.0\pm0.5$&$14.0\pm0.5$&$16.3\pm0.5$\\
8&$8.4\pm0.4$&$7.9\pm0.3$&$9.2\pm0.4$\\
9&$18.4\pm0.4$&$17.3\pm0.3$&$20.1\pm0.5$\\
10&$5.6\pm0.4$&$5.2\pm0.3$&$6.1\pm0.5$\\
12&$13.7\pm0.4$&$12.9\pm0.3$&$15.0\pm0.5$\\
G351.62&$5.2\pm0.2$&$4.9\pm0.2$&$5.7\pm0.3$\\
&$1.0\pm0.1$&$0.9\pm0.1$&$1.1\pm0.1$\\
\hline
\end{tabular}
\end{center}
\end{table}

\subsection{{\rm CO}--${\rm N_2H^+}$ enhancement}\label{co_n2h+}

For reasons not yet fully understood, ${\rm N_2H^+}$ (along with
ammonia), unlike CO, is resistant to depletion in very dense and cold
cores \citep{2005ApJ...621L..33O}. However, internal heating
re-introduces CO to the gas phase.  Since CO is one of the
main destroyers of ${\rm N_2H^+}$ in the gas phase \citep[e.g., ][]{2001ApJ...552..639A}, the abundance
of N$_2$H$^+$ decreases when protostellar/cluster formation  heats up the envelope
to sufficiently high temperatures, releasing the frozen out CO.
The ratio of the observed CO-to-${\rm N_2H^+}$ abundance should
thus reflect this chemistry. Hence, this differential depletion of the
two species might therefore be used as a chemical clock to assess the evolutionary status of the
clumps along the filament.

After smoothing the LABOCA data to the same resolution as the
${\rm N_2H^+}$ and C$^{17}$O data, we estimated the molecular
abundances of the two species and the relative abundance of C$^{17}$O to N$_2$H$^+$. Results are
tabulated in Table~\ref{tab_n2h+}.  For this, we used the C$^{17}$O column
density assuming an
excitation temperature of 20~K. Based on the absence/presence of MIR
emission, hence on stellar content, we expected a significant spread in relative
 abundance between the clumps. Indeed, the ${\rm N_2H^+}$ column density
correlates well with the MIR brightness with the column density
increasing from clumps 5  to  1.

The ratio of ${\rm C^{17}O}$ and ${\rm
   N_2H^+}$ abundance is visualised in
 Fig.~\ref{fig_n2h+_dust_co}.  To do this, we have taken  the CO and N$_2$H$^+$ abundance uncertainties into account. The
   statistical uncertainties based on fits to the line profile is
   sufficient for N$_2$H$^+$. However for CO, while the errors on the
   fit to line profile is only a few percent, the unknown  CO excitation temperature,
   $T_{ex_{\rm{CO}}}$, makes it the most significant parameter for
   estimating the uncertainties. We therefore vary the excitation
   temperature between 15 and 25~K and refer to Table~\ref{cd} for
   extremes in ${\rm C^{17}O}$ column densities. For the densities observed in the
   cloud ($> 10^{5}$~~cm$^{3}$), $T_{ex_{\rm{CO}}}$ must be very close
   to the gas temperature. Based on ammonia (1,1) and (2,2)
   observations with the Parkes telescope towards clumps 1, 2, and 3, Wienen et al. (in
   prep.) find a temperature of $<$20~K. For clumps 2 and 5, we thus adopt a reasonable temperature range
   of 10~K to 25~K to calculate the column densities.  The mean of the largest difference in ${\rm C^{17}O}$
   column densities at these very different temperatures
   (Table~\ref{cd}) is then considered as a conservative upper limit
   for the error on the ratio. A higher
   temperature extreme of 35~K is adopted for clump 1 which hosts the
   hot core associated with IRAS\,17233--3606.

Uncertainties in
     the beam filling factor of N$_2$H$^+$ and C$^{17}$O is an
     additional source of error. For the C$^{17}$O abundance, we expect a value close to 1 for the C$^{17}$O column density 
(because of the low critical density of the line), and also for the H$_2$ column density given the size estimates obtained 
in Sect.~\ref{870}; however, since we are limited by our single
pointing measurements for CO and poor resolution for N$_2$H$^+$, the
respective beam filling factor remains unknown. For the
N$_2$H$^+$ abundance, dust emission and N$_2$H$^+$ have very similar
distributions. Also, the significant differences in source sizes
extracted from our higher resolution dust
observations (see Table~\ref{clfn}) indicate that beam dilution might play a role.
     The abundance
 ratio decreases from 437/214 (clump 1) to 93 for clump 5 from  the
 brightest end of the filament to the other end; however, based on the
 uncertainties discussed above, the observed differences in abundance
 ratio cannot be ascertained.  The filament has an average ${\rm
N_2H^+}$ abundance (over the three clumps) of 2.0$\times 10^{-10}$
relative to H$_2$. The derived column density and abundance is
consistent with those found in other infrared dark clouds
(\citealt{2006ApJS..166..567R,2008ApJ...678.1049S}).

\begin{table*}
\begin{center}
\caption{ {${\rm{N_2H^{+}}}$} 1--0 line parameters.}\label{tab_n2h+}
\begin{tabular}{lrcrrcrccc}
\hline
\hline

\multicolumn{1}{c}{Clump} &\multicolumn{1}{c}{$\rm{v}_{\rm
   LSR}$}&\multicolumn{1}{c}{$\Delta
 \rm{v}$}&\multicolumn{1}{c}{$\tau_{\rm tot}$}&\multicolumn{1}{c}{$T_{\rm   EX}$}&\multicolumn{1}{c}{$N_{\rm{N_2H^{+}}}$}&\multicolumn{1}{c}{$\chi_{\rm{N_2H^{+}}}$}&\multicolumn{1}{c}{$\chi_{\rm{C^{17}O}}$} & $\chi_{\rm{C^{17}O}}$/$\chi_{\rm{N_{2}H^+}}$ \\
\multicolumn{1}{c}{}      &\multicolumn{1}{c}{[km~s$^{-1}$]}&\multicolumn{1}{c}{[km~s$^{-1}$]}&\multicolumn{1}{c}{}& \multicolumn{1}{c}{[K]}&\multicolumn{1}{c}{[10$^{13}$~cm$^{-2}$]}&  \multicolumn{1}{c}{[10$^{-10}$]}      &     {[10$^{-8}$]}        &                          \\

\hline

1 &  -3.7  (0.04)  & 4.55  (0.06) &   0.2 - 1 &    11.7 - 37.3 &
5.5 - 11.3  & 0.83 - 1.68 & 3.6 & 214 - 437 \\
2 &  -2.9  (0.01)  & 3.06  (0.05) &  1.05  (0.02) &     9.2     (0.4)
&  2.5 (0.2)   & 2.3  & 7.1 & 307 \\
5 &  -2.2  (0.04)  & 3.36  (0.09) &  1.41  (0.40) &     6.1     (1.1) &  1.8 (0.8)    & 3.7  & 3.4 & 93 \\
\hline

\end{tabular}
\end{center}
Note $\tau_{\rm tot}$ is the sum of the peak optical depth of the seven hyperfine components, $\chi_{\rm{N_2H^{+}}}$ ($\chi_{\rm{C^{17}O}}$) is
the N$_2$H$^+$ (${\rm{C^{17}O}}$) abundance. \\
\end{table*}

The fairly constant CO abundance over clumps with very distinct levels of star formation suggests that processes other than desorption of CO might affect the ${\rm N_2H^+}$
  chemistry. In a CS and ${\rm N_2H^+}$ study of a small sample of
  dense cores in high-mass star-forming regions,
  \citet{2007A&A...461..523P} briefly explore alternate pathways for
  the dissociative recombination of ${\rm N_2H^+}$ other than the
  standard channel with $\rm N_2$ as the end product (\citealt{2004ApJ...609..459G}).  However, these pathways may not be dominant, as previously thought (\citealt{2009JPhCS.192a2004A}), and new chemical models are required to explain this. Observationally,  HCO$^+$/H$^{13}$CO$^+$ distribution and abundance might help us in fully ruling out the role of CO desorption.

\subsection{\rm{N$_2$}H$^+$ line width and velocity field}
\begin{figure*}
\centering \includegraphics[angle=-90,width=19cm]{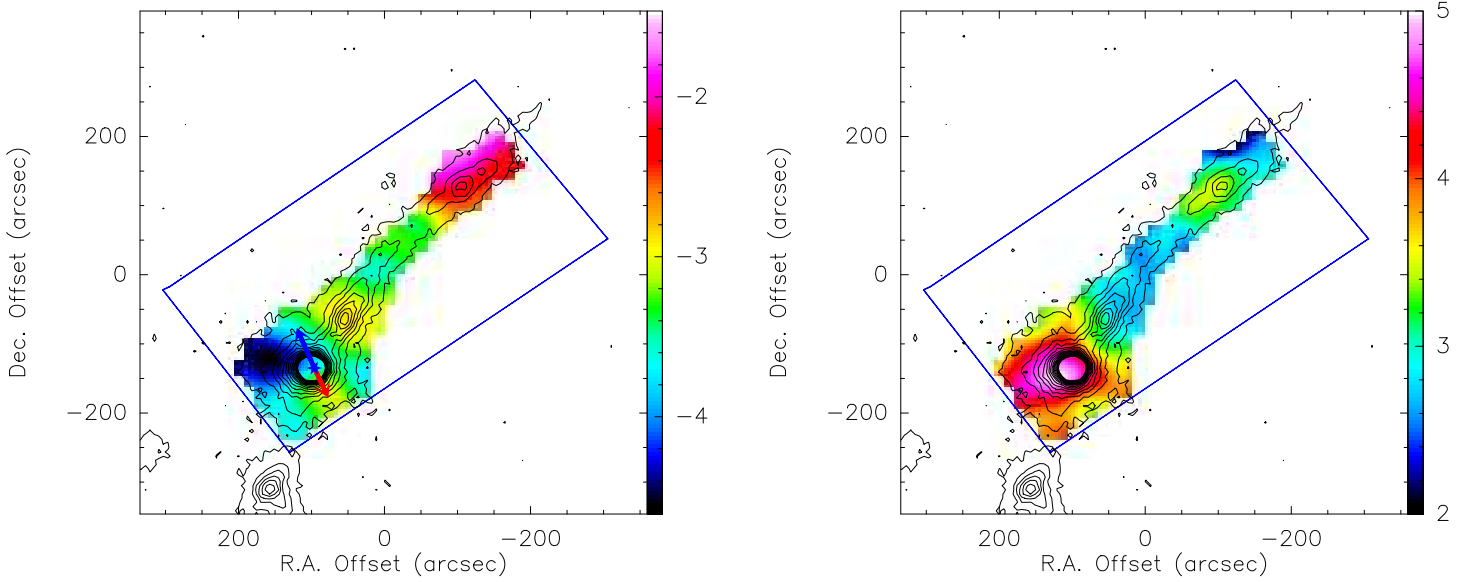}
\caption{{\it Left Panel}: MOPRA N$_2$H$^+$ velocity distribution with dust continuum LABOCA 870$\mu$m
emission in contours. Red and blue arrows indicate the rough direction
of the red and blue CO outflow lobes detected towards IRAS\,17233--3606 respectively. {\it Right Panel}:
MOPRA N$_2$H$^+$ line width distribution ({\it FWHM}) with dust continuum
LABOCA 870$\mu$m emission in contours. In both maps, background image
units are in ${\rm kms^{-1}}$, and only those pixels that correspond to
N$_2$H$^+$ integrated intensity ${\rm > 2.4 Jy~kms^{-1}}$ have been
considered significant.}\label{fig_n2h+_parms}
\end{figure*}

We have analysed the line width and velocity fields from the
N$_2$H$^+$ 3-D data, and the results are shown in
Fig.~\ref{fig_n2h+_parms}. For these, we have considered only pixels
corresponding to N$_2$H$^+$ integrated intensity $> 3\sigma$.

Four
clumps are identified in the first moment map (see Fig.~\ref{fig_n2h+_parms}, left panel). 
These have small differences in velocities 
compared to their line widths (Fig.~\ref{fig_n2h+_parms}, right panel).
The velocity distribution for clump 1 is likely complicated by the intense star formation activity.
If the filament were to be part of a shell-like structure (or if it were rotating), one would expect a clear linear velocity gradient from one end of the filament to the other. Most of the extended gas appears to be at a velocity of $-3.4~$kms$^{-1}$, which we take to be  the velocity of the ambient medium. Clumps 2 and 5  are separated by this lower density extended medium, contrary to a linear velocity gradient (see Sect.~\ref{disc}). 
A projection effect where clumps 2 and  5 are coincidentally
aligned along a ring also seems unlikely given that there is no obvious kink in the velocity field.

Of particular interest are two striking features in the velocity map; (i) a large-scale
($\approx 1$~parsec) NE--SW velocity gradient of more than 1.6~${\rm
kms^{-1}}$ associated with the hot core in IRAS\,17233--3606 (clump 1), and (ii) a smaller scale velocity gradient
 ($\approx 0.6$~parsec) in the northern clump (clumps 5, 6). Both velocity
gradients are perpendicular to the filament, but in different
directions. Interestingly, the NE(blue)-SW(red) gradient towards the hot core correlates
well with the direction and kinematics of the CO outflows \citep{2008A&A...485..167L,2009A&A...507.1443L}.
Although N$_2$H$^+$ is seldom associated
with outflows, the interactions of outflow shocks with the molecular
envelope (traced by N$_2$H$^+$) could explain such a gradient towards
the hot core (clump 1) as detected in the same transition  
towards the class 0 object IRAM\,04191+1522 by \citet{2005ApJ...619..948L}.
The axis of the outflow (shown in Fig.~\ref{fig_n2h+_parms}) would favour this shock scenario.  This would also explain 
the very broad lines
($\approx 5$~${\rm kms^{-1}}$) observed towards the hot core.  Such
broadening away from the mm core has also been observed in lower excitation NH$_3$ lines in
low- and high-mass star-forming regions \citep{1993ApJ...417L..45B,1999ApJ...527L.117Z,2007A&A...466..989B}.
A large-scale infall motion would also account for the observed velocity gradient. This large-scale infall is often observed in regions of high-mass star formation \citep{2006MNRAS.367..553P}.

The velocities at the peak positions between the northern and southern
clumps differ by at least $> 1.6$~${\rm kms^{-1}}$. This is comparable
to the average velocity dispersion (see
Fig.~\ref{fig_n2h+_parms} right panel, and Table~\ref{tab_n2h+}) of the clumps
within the filament. At this rate, the clumps would separate out on a
scale of 1.6~pc in 1~Myrs.  

The line width distribution shown in
Fig.~\ref{fig_n2h+_parms} might reflect differences in the evolutionary status of the
clumps along the filament. As expected, the hot core (clump 1)
exhibits the broadest line, and the rest of the filament (clumps 2, 5, and the rest of the cloud) has roughly similar line widths with a slightly higher peak
towards clump 5. Clump 5 is associated with a stronger 8$\mu$m
emission than clump 2; however, in all cases the
line width is highly supersonic.  

Given the line width and the extent of the cloud in projection,
  we can derive the crossing time.  Thus the observed turbulents motions within the
  arc of filament would provide an upper limit to its lifetime of less than 1 Myr.

 Unlike CO and its
 isotopologues studied here, the N$_2$H$^+$ transition has a high
 critical density ($> 10^{5}$~~cm$^{3}$). Therefore, in spite of the
 lower resolution of the  N$_2$H$^+$ data, we are selectively probing the dense
 gas, and not the envelope; i.e., line
 width-size relation does not necessarily apply. This might have the important
implication that, although local star formation might be the cause of
the exceptional line broadening in clumps 1 and 5, an external agent might be the
cause of the line broadening along the whole filament.   Interestingly, evidence of large-scale shocks, maybe  
remnants of the IRDC 
formation process,
were found by \citet{2010MNRAS.406..187J} towards IRDC G035.39--00.33.
The
 present data set is insufficient to properly address this
 scenario. Future higher angular resolution observations capable of resolving the dense cores structures will allow us to test this scenario.

\section{Discussion \label{disc}}
Recently, \citet{2010ApJ...715..310R} have studied a sample of 38 IRDCs in continuum emission
at submillimetre and mid-- to far--IR wavelengths. They found that only a fifth
of their clumps have 8$\mu$m emission, while almost half of them have a 24$\mu$m point source.
In the case of G351.77, the majority of identified dust clumps are associated with emission at 24$\mu$m, suggesting
that  star formation is going on along the full extent of the filament.
This is also confirmed by H$_2$ observations at 2.12$\mu$m which  detected
weak outflows  along the full extent of the filament
(Stanke et al. in prep.). The higher detection rate of 24$\mu$m compact sources in G351.77 compared
to the sample studied by \citet{2010ApJ...715..310R} could be due to higher sensitivity to low-luminosity objects, given
the closer distance of G351.77 ($D\le 1$~kpc) compared to the typical distance ($\sim$4~kpc) of the IRDCs of
\citet{2010ApJ...715..310R}.

The velocity structure detected in N$_2$H$^+$ shows very strong
gradients towards individual clumps. This might indicate a general infall
within the individual clumps (which might harbour a protocluster) or
shocks in molecular envelopes.  The line widths are highly supersonic
and might also indicate the presence of such shocks.  

The 8$\mu$m map (Fig.~\ref{ring}) shows emission along a semi-shell like structure,
which is reminiscent of bubbles of gas and dust detected around
expanding H{\sc ii} regions \citep[e.g.,][]{1998ASPC..148..150E,2003A&A...408L..25D,2005A&A...433..565D,2010A&A...523A...6D}.
These structures are potential candidates
of star formation triggered by massive stars.
The continuum emission at 870$\mu$m is also shaped along a semicircular structure.
Therefore a tempting explanation for the broad line profiles associated with
dust clumps in G351.77 is shock compression due to expansion of H{\sc ii}
regions on pre-existing gas. This could also be the triggering event  for the
intense star formation activity shown by several molecular
outflows along the whole G351.77 (Stanke et
al. in prep).
We searched the SIMBAD\footnote{This research has made use of the SIMBAD database,
operated at CDS, Strasbourg, France} database for all
astronomical objects contained in a circle of 0.5$^\circ$ diameter with
clump 7 (in the middle of the dark filament)  at
its centre, but could not identify any obvious source of trigger.
In particular, the only cm continuum sources in the region found in the NVSS catalogue \citep{1998AJ....115.1693C}
are associated with
IRAS\,17233--3606 and  IRAS\,17221--3619.
From the velocity field derived from
the CO observations, we know that the LABOCA semicircular structure is
associated with molecular gas at two different velocities, one at
 $\sim -3$~km~s$^{-1}$ (IRAS\,17233--3606, 
G351.77 and clump 3), the other at $\sim -22$~km~s$^{-1}$ (IRAS\,17221--3619, IRAS\,17227--3619 and the position G351.62).
The near kinematic distance of  IRAS\,17221--3619 and IRAS\,17227--3619 is $\sim 3.6$~kpc ($D_{\rm FAR}\sim 13$~kpc), while the
 two velocity components
of the position G351.62   correspond to kinematic distances of $\sim 600$~pc and $\sim 3.6$~kpc
($D_{\rm FAR}\sim 16$ and 13~kpc, respectively), derived by using the rotation curve of \citet{1993A&A...275...67B}.
For IRAS\,17227--3619, \citet{2002A&A...381..571P} derived a kinematic distance of 3.4~kpc based on IR recombination lines, while \citet{2006A&A...455..923B}
estimated distances between 1.9 and 3.2~kpc based on H- and K-band spectra of the source.
Thus,   the semicircular structure detected at 870$\mu$m in the
LABOCA map is likely  a projection effect of two different molecular clouds.
However, from our pointed CO observations we can neither infer the true
distribution of the two cloud complexes nor  verify whether  the
 western half circle, delineated by
diffuse 8$\mu$m emission (see Sect.~\ref{ir}), is connected to one of these two clouds.

\section{Conclusions}
In summary, we have characterised the G351.77 dense filament in thermal dust
continuum, N$_2$H$^+$ $(1-0)$ and CO (various isotopologues at low
excitation) emission.  The molecular line emission traces the thermal dust
continuum very well. Three main clumps in N$_2$H$^+$ and eight clumps
(at higher resolution) in the dust continuum were observed. These clumps
 also have associated emission from CO isotopologues. We  analysed the integrated
intensity, velocity, and line width distribution along the
filament. The column density and mass of the clumps were also 
derived. The velocity structure shows very strong gradients across
two clumps (1 and 5). This might indicate either a global infall
within the individual clumps (which might harbour a protocluster) or
shocks in molecular envelopes due to outflow-envelope interaction.  All the clumps have supersonic line
widths. Such supersonic line widths along the filament might also indicate the
presence of shocks.  Clearly, the most evolved clump is the hot core
that shows (i) the brightest MIR emission, (ii) broadest line width,
(iii) largest velocity gradient, and the (iv) most massive and
dense core. Most mm clumps have a mass $>10$~$M_\odot$. 

Our study shows that  G351.77  has ongoing star formation in different evolutionary stages along the filament.
Therefore, given the relatively close distance of this object compared to typical sites of massive star formation,
G351.77 is an ideal target  
for investigating the role of IRDCs in the early phases of the formation of massive stars and stellar clusters, and
for studying the population of low-mass stars in these objects.

\begin{acknowledgements}
The authors would like to thank J. Kauffmann for fruitful discussions.
S.T. is grateful to the Deutsche Forschungsgemeinschaft for a research
grant (TH1301/3-1). T.P. acknowledges support from the Combined Array for Research in Millimeter-wave Astronomy (CARMA), which is supported by the National Science Foundation through grant AST 05-40399.
\end{acknowledgements}


\begin{thebibliography}{82}
\expandafter\ifx\csname natexlab\endcsname\relax\def\natexlab#1{#1}\fi

\bibitem[{{Adams} {et~al.}(2009){Adams}, {Molek}, \&
  {McLain}}]{2009JPhCS.192a2004A}
{Adams}, N.~G., {Molek}, C.~D., \& {McLain}, J.~L. 2009, Journal of Physics
  Conference Series, 192, 012004

\bibitem[{{Aikawa} {et~al.}(2001){Aikawa}, {Ohashi}, {Inutsuka}, {Herbst}, \&
  {Takakuwa}}]{2001ApJ...552..639A}
{Aikawa}, Y., {Ohashi}, N., {Inutsuka}, S., {Herbst}, E., \& {Takakuwa}, S.
  2001, \apj, 552, 639

\bibitem[{{Andr{\'e}} {et~al.}(2007){Andr{\'e}}, {Belloche}, {Motte}, \&
  {Peretto}}]{2007A&A...472..519A}
{Andr{\'e}}, P., {Belloche}, A., {Motte}, F., \& {Peretto}, N. 2007, \aap, 472,
  519

\bibitem[{{Bachiller} {et~al.}(1993){Bachiller}, {Martin-Pintado}, \&
  {Fuente}}]{1993ApJ...417L..45B}
{Bachiller}, R., {Martin-Pintado}, J., \& {Fuente}, A. 1993, \apjl, 417, L45

\bibitem[{{Belloche} {et~al.}(2011){Belloche}, {Schuller}, {Parise},
  {Andr{\'e}}, {Hatchell}, {J{\o}rgensen}, {Bontemps}, {Wei{\ss}}, {Menten}, \&
  {Muders}}]{2011A&A...527A.145B}
{Belloche}, A., {Schuller}, F., {Parise}, B., {et~al.} 2011, \aap, 527, A145+

\bibitem[{{Benjamin} {et~al.}(2003){Benjamin}, {Churchwell}, {Babler}, {Bania},
  {Clemens}, {Cohen}, {Dickey}, {Indebetouw}, {Jackson}, {Kobulnicky},
  {Lazarian}, {Marston}, {Mathis}, {Meade}, {Seager}, {Stolovy}, {Watson},
  {Whitney}, {Wolff}, \& {Wolfire}}]{2003PASP..115..953B}
{Benjamin}, R.~A., {Churchwell}, E., {Babler}, B.~L., {et~al.} 2003, \pasp,
  115, 953

\bibitem[{{Bergin} {et~al.}(2002){Bergin}, {Alves}, {Huard}, \&
  {Lada}}]{2002ApJ...570L.101B}
{Bergin}, E.~A., {Alves}, J., {Huard}, T., \& {Lada}, C.~J. 2002, \apjl, 570,
  L101

\bibitem[{{Beuther} \& {Schilke}(2004)}]{2004science}
{Beuther}, H. \& {Schilke}, P. 2004, Science, 303, 1167

\bibitem[{{Beuther} {et~al.}(2007){Beuther}, {Walsh}, {Thorwirth}, {Zhang},
  {Hunter}, {Megeath}, \& {Menten}}]{2007A&A...466..989B}
{Beuther}, H., {Walsh}, A.~J., {Thorwirth}, S., {et~al.} 2007, \aap, 466, 989

\bibitem[{{Borissova} {et~al.}(2006){Borissova}, {Ivanov}, {Minniti}, \&
  {Geisler}}]{2006A&A...455..923B}
{Borissova}, J., {Ivanov}, V.~D., {Minniti}, D., \& {Geisler}, D. 2006, \aap,
  455, 923

\bibitem[{{Brand} \& {Blitz}(1993)}]{1993A&A...275...67B}
{Brand}, J. \& {Blitz}, L. 1993, \aap, 275, 67

\bibitem[{{Bronfman} {et~al.}(1996){Bronfman}, {Nyman}, \&
  {May}}]{1996A&AS..115...81B}
{Bronfman}, L., {Nyman}, L.-A., \& {May}, J. 1996, \aaps, 115, 81

\bibitem[{{Carey} {et~al.}(1998){Carey}, {Clark}, {Egan}, {Price}, {Shipman},
  \& {Kuchar}}]{1998ApJ...508..721C}
{Carey}, S.~J., {Clark}, F.~O., {Egan}, M.~P., {et~al.} 1998, \apj, 508, 721

\bibitem[{{Carey} {et~al.}(2000){Carey}, {Feldman}, {Redman}, {Egan},
  {MacLeod}, \& {Price}}]{2000ApJ...543L.157C}
{Carey}, S.~J., {Feldman}, P.~A., {Redman}, R.~O., {et~al.} 2000, \apjl, 543,
  L157

\bibitem[{{Carey} {et~al.}(2009){Carey}, {Noriega-Crespo}, {Mizuno}, {Shenoy},
  {Paladini}, {Kraemer}, {Price}, {Flagey}, {Ryan}, {Ingalls}, {Kuchar},
  {Pinheiro Gon{\c c}alves}, {Indebetouw}, {Billot}, {Marleau}, {Padgett},
  {Rebull}, {Bressert}, {Ali}, {Molinari}, {Martin}, {Berriman}, {Boulanger},
  {Latter}, {Miville-Deschenes}, {Shipman}, \& {Testi}}]{2009PASP..121...76C}
{Carey}, S.~J., {Noriega-Crespo}, A., {Mizuno}, D.~R., {et~al.} 2009, \pasp,
  121, 76

\bibitem[{{Caselli} {et~al.}(2002){Caselli}, {Walmsley}, {Zucconi}, {Tafalla},
  {Dore}, \& {Myers}}]{2002ApJ...565..344C}
{Caselli}, P., {Walmsley}, C.~M., {Zucconi}, A., {et~al.} 2002, \apj, 565, 344

\bibitem[{{Caswell} {et~al.}(1980){Caswell}, {Haynes}, \&
  {Phys}}]{1980IAUC.3509....2C}
{Caswell}, J.~L., {Haynes}, R.~F., \& {Phys}, J. 1980, \iaucirc, 3509, 2

\bibitem[{{Churchwell} {et~al.}(2007){Churchwell}, {Watson}, {Povich},
  {Taylor}, {Babler}, {Meade}, {Benjamin}, {Indebetouw}, \&
  {Whitney}}]{2007ApJ...670..428C}
{Churchwell}, E., {Watson}, D.~F., {Povich}, M.~S., {et~al.} 2007, \apj, 670,
  428

\bibitem[{{Condon} {et~al.}(1998){Condon}, {Cotton}, {Greisen}, {Yin},
  {Perley}, {Taylor}, \& {Broderick}}]{1998AJ....115.1693C}
{Condon}, J.~J., {Cotton}, W.~D., {Greisen}, E.~W., {et~al.} 1998, \aj, 115,
  1693

\bibitem[{{de Wit} {et~al.}(2004){de Wit}, {Testi}, {Palla}, {Vanzi}, \&
  {Zinnecker}}]{2004A&A...425..937D}
{de Wit}, W.~J., {Testi}, L., {Palla}, F., {Vanzi}, L., \& {Zinnecker}, H.
  2004, \aap, 425, 937

\bibitem[{{Deharveng} {et~al.}(2003){Deharveng}, {Lefloch}, {Zavagno},
  {Caplan}, {Whitworth}, {Nadeau}, \& {Mart{\'{\i}}n}}]{2003A&A...408L..25D}
{Deharveng}, L., {Lefloch}, B., {Zavagno}, A., {et~al.} 2003, \aap, 408, L25

\bibitem[{{Deharveng} {et~al.}(2010){Deharveng}, {Schuller}, {Anderson},
  {Zavagno}, {Wyrowski}, {Menten}, {Bronfman}, {Testi}, {Walmsley}, \&
  {Wienen}}]{2010A&A...523A...6D}
{Deharveng}, L., {Schuller}, F., {Anderson}, L.~D., {et~al.} 2010, \aap, 523,
  A6

\bibitem[{{Deharveng} {et~al.}(2005){Deharveng}, {Zavagno}, \&
  {Caplan}}]{2005A&A...433..565D}
{Deharveng}, L., {Zavagno}, A., \& {Caplan}, J. 2005, \aap, 433, 565

\bibitem[{{Egan} {et~al.}(1998){Egan}, {Shipman}, {Price}, {Carey}, {Clark}, \&
  {Cohen}}]{1998ApJ...494L.199E}
{Egan}, M.~P., {Shipman}, R.~F., {Price}, S.~D., {et~al.} 1998, \apjl, 494,
  L199

\bibitem[{{Elmegreen}(1998)}]{1998ASPC..148..150E}
{Elmegreen}, B.~G. 1998, in Astronomical Society of the Pacific Conference
  Series, Vol. 148, Origins, ed. {C.~E.~Woodward, J.~M.~Shull, \&
  H.~A.~Thronson Jr.}, 150

\bibitem[{{Fa{\'u}ndez} {et~al.}(2004){Fa{\'u}ndez}, {Bronfman}, {Garay},
  {Chini}, {Nyman}, \& {May}}]{2004A&A...426...97F}
{Fa{\'u}ndez}, S., {Bronfman}, L., {Garay}, G., {et~al.} 2004, \aap, 426, 97

\bibitem[{{Fix} {et~al.}(1982){Fix}, {Mutel}, {Gaume}, \&
  {Claussen}}]{1982ApJ...259..657F}
{Fix}, J.~D., {Mutel}, R.~L., {Gaume}, R.~A., \& {Claussen}, M.~J. 1982, \apj,
  259, 657

\bibitem[{{Geppert} {et~al.}(2004){Geppert}, {Thomas}, {Semaniak}, {Ehlerding},
  {Millar}, {{\"O}sterdahl}, {af Ugglas}, {Djuri{\'c}}, {Pa{\'a}l}, \&
  {Larsson}}]{2004ApJ...609..459G}
{Geppert}, W.~D., {Thomas}, R., {Semaniak}, J., {et~al.} 2004, \apj, 609, 459

\bibitem[{{Goodman} {et~al.}(2009){Goodman}, {Rosolowsky}, {Borkin}, {Foster},
  {Halle}, {Kauffmann}, \& {Pineda}}]{2009Natur.457...63G}
{Goodman}, A.~A., {Rosolowsky}, E.~W., {Borkin}, M.~A., {et~al.} 2009, \nat,
  457, 63

\bibitem[{{G{\"u}sten} {et~al.}(2006){G{\"u}sten}, {Nyman}, {Schilke},
  {Menten}, {Cesarsky}, \& {Booth}}]{2006A&A...454L..13G}
{G{\"u}sten}, R., {Nyman}, L.~{\AA}., {Schilke}, P., {et~al.} 2006, \aap, 454,
  L13

\bibitem[{{Hernandez} \& {Tan}(2011)}]{2011ApJ...730...44H}
{Hernandez}, A.~K. \& {Tan}, J.~C. 2011, \apj, 730, 44

\bibitem[{{Hofner} {et~al.}(2000){Hofner}, {Wyrowski}, {Walmsley}, \&
  {Churchwell}}]{2000ApJ...536..393H}
{Hofner}, P., {Wyrowski}, F., {Walmsley}, C.~M., \& {Churchwell}, E. 2000,
  \apj, 536, 393

\bibitem[{{Jim{\'e}nez-Serra} {et~al.}(2010){Jim{\'e}nez-Serra}, {Caselli},
  {Tan}, {Hernandez}, {Fontani}, {Butler}, \& {van Loo}}]{2010MNRAS.406..187J}
{Jim{\'e}nez-Serra}, I., {Caselli}, P., {Tan}, J.~C., {et~al.} 2010, \mnras,
  406, 187

\bibitem[{{Johnstone} {et~al.}(2004){Johnstone}, {Di Francesco}, \&
  {Kirk}}]{2004ApJ...611L..45J}
{Johnstone}, D., {Di Francesco}, J., \& {Kirk}, H. 2004, \apjl, 611, L45

\bibitem[{{Johnstone} {et~al.}(2010){Johnstone}, {Rosolowsky}, {Tafalla}, \&
  {Kirk}}]{2010ApJ...711..655J}
{Johnstone}, D., {Rosolowsky}, E., {Tafalla}, M., \& {Kirk}, H. 2010, \apj,
  711, 655

\bibitem[{{Kauffmann} {et~al.}(2008){Kauffmann}, {Bertoldi}, {Bourke}, {Evans},
  \& {Lee}}]{2008A&A...487..993K}
{Kauffmann}, J., {Bertoldi}, F., {Bourke}, T.~L., {Evans}, II, N.~J., \& {Lee},
  C.~W. 2008, \aap, 487, 993

\bibitem[{{Klein} {et~al.}(2006){Klein}, {Philipp}, {Kr{\"a}mer}, {Kasemann},
  {G{\"u}sten}, \& {Menten}}]{2006A&A...454L..29K}
{Klein}, B., {Philipp}, S.~D., {Kr{\"a}mer}, I., {et~al.} 2006, \aap, 454, L29

\bibitem[{{Lada} \& {Lada}(2003)}]{2003ARA&A..41...57L}
{Lada}, C.~J. \& {Lada}, E.~A. 2003, \araa, 41, 57

\bibitem[{{Ladd} {et~al.}(2005){Ladd}, {Purcell}, {Wong}, \&
  {Robertson}}]{2005PASA...22...62L}
{Ladd}, N., {Purcell}, C., {Wong}, T., \& {Robertson}, S. 2005, \pasa, 22, 62

\bibitem[{{Lee} {et~al.}(2005){Lee}, {Ho}, \& {White}}]{2005ApJ...619..948L}
{Lee}, C., {Ho}, P.~T.~P., \& {White}, S.~M. 2005, \apj, 619, 948

\bibitem[{{Leurini} {et~al.}(2011){Leurini}, {Codella}, {Zapata}, {Beltran},
  {Schilke}, \& {Cesaroni}}]{2011arXiv1104.0857L}
{Leurini}, S., {Codella}, C., {Zapata}, L., {et~al.} 2011, \aap

\bibitem[{{Leurini} {et~al.}(2009){Leurini}, {Codella}, {Zapata}, {Belloche},
  {Stanke}, {Wyrowski}, {Schilke}, {Menten}, \&
  {G{\"u}sten}}]{2009A&A...507.1443L}
{Leurini}, S., {Codella}, C., {Zapata}, L.~A., {et~al.} 2009, \aap, 507, 1443

\bibitem[{{Leurini} {et~al.}(2008){Leurini}, {Hieret}, {Thorwirth}, {Wyrowski},
  {Schilke}, {Menten}, {G{\"u}sten}, \& {Zapata}}]{2008A&A...485..167L}
{Leurini}, S., {Hieret}, C., {Thorwirth}, S., {et~al.} 2008, \aap, 485, 167

\bibitem[{{M{\" u}ller} {et~al.}(2001){M{\" u}ller}, {Thorwirth}, {Roth}, \&
  {Winnewisser}}]{2001A&A...370L..49M}
{M{\" u}ller}, H.~S.~P., {Thorwirth}, S., {Roth}, D.~A., \& {Winnewisser}, G.
  2001, \aap, 370, L49

\bibitem[{{Mart{\'{\i}}n-Hern{\'a}ndez}
  {et~al.}(2003){Mart{\'{\i}}n-Hern{\'a}ndez}, {van der Hulst}, \&
  {Tielens}}]{2003A&A...407..957M}
{Mart{\'{\i}}n-Hern{\'a}ndez}, N.~L., {van der Hulst}, J.~M., \& {Tielens},
  A.~G.~G.~M. 2003, \aap, 407, 957

\bibitem[{{Menten}(1991)}]{1991ApJ...380L..75M}
{Menten}, K.~M. 1991, \apjl, 380, L75

\bibitem[{{Miettinen} {et~al.}(2006){Miettinen}, {Harju}, {Haikala}, \&
  {Pomr{\'e}n}}]{2006A&A...460..721M}
{Miettinen}, O., {Harju}, J., {Haikala}, L.~K., \& {Pomr{\'e}n}, C. 2006, \aap,
  460, 721

\bibitem[{{Molinari} {et~al.}(1998){Molinari}, {Testi}, {Brand}, {Cesaroni}, \&
  {Palla}}]{1998ApJ...505L..39M}
{Molinari}, S., {Testi}, L., {Brand}, J., {Cesaroni}, R., \& {Palla}, F. 1998,
  \apjl, 505, L39

\bibitem[{{Molinari} {et~al.}(2002){Molinari}, {Testi}, {Rodr{\'{\i}}guez}, \&
  {Zhang}}]{2002ApJ...570..758M}
{Molinari}, S., {Testi}, L., {Rodr{\'{\i}}guez}, L.~F., \& {Zhang}, Q. 2002,
  \apj, 570, 758

\bibitem[{{Motte} {et~al.}(1998){Motte}, {Andre}, \&
  {Neri}}]{1998A&A...336..150M}
{Motte}, F., {Andre}, P., \& {Neri}, R. 1998, \aap, 336, 150

\bibitem[{{Motte} {et~al.}(2007){Motte}, {Bontemps}, {Schilke}, {Schneider},
  {Menten}, \& {Brogui{\`e}re}}]{2007A&A...476.1243M}
{Motte}, F., {Bontemps}, S., {Schilke}, P., {et~al.} 2007, \aap, 476, 1243

\bibitem[{{Myers}(2009)}]{2009ApJ...700.1609M}
{Myers}, P.~C. 2009, \apj, 700, 1609

\bibitem[{{{\"O}berg} {et~al.}(2005){{\"O}berg}, {van Broekhuizen}, {Fraser},
  {Bisschop}, {van Dishoeck}, \& {Schlemmer}}]{2005ApJ...621L..33O}
{{\"O}berg}, K.~I., {van Broekhuizen}, F., {Fraser}, H.~J., {et~al.} 2005,
  \apjl, 621, L33

\bibitem[{{Olmi} \& {Testi}(2002)}]{2002A&A...392.1053O}
{Olmi}, L. \& {Testi}, L. 2002, \aap, 392, 1053

\bibitem[{{Ossenkopf} \& {Henning}(1994)}]{1994A&A...291..943O}
{Ossenkopf}, V. \& {Henning}, T. 1994, \aap, 291, 943

\bibitem[{{Peeters} {et~al.}(2002){Peeters}, {Mart{\'{\i}}n-Hern{\'a}ndez},
  {Damour}, {Cox}, {Roelfsema}, {Baluteau}, {Tielens}, {Churchwell}, {Kessler},
  {Mathis}, {Morisset}, \& {Schaerer}}]{2002A&A...381..571P}
{Peeters}, E., {Mart{\'{\i}}n-Hern{\'a}ndez}, N.~L., {Damour}, F., {et~al.}
  2002, \aap, 381, 571

\bibitem[{{Penzias}(1981)}]{1981ApJ...249..518P}
{Penzias}, A.~A. 1981, \apj, 249, 518

\bibitem[{{Perault} {et~al.}(1996){Perault}, {Omont}, {Simon}, {Seguin},
  {Ojha}, {Blommaert}, {Felli}, {Gilmore}, {Guglielmo}, {Habing}, {Price},
  {Robin}, {de Batz}, {Cesarsky}, {Elbaz}, {Epchtein}, {Fouque}, {Guest},
  {Levine}, {Pollock}, {Prusti}, {Siebenmorgen}, {Testi}, \&
  {Tiphene}}]{1996A&A...315L.165P}
{Perault}, M., {Omont}, A., {Simon}, G., {et~al.} 1996, \aap, 315, L165

\bibitem[{{Pillai} {et~al.}(2006{\natexlab{a}}){Pillai}, {Wyrowski}, {Carey},
  \& {Menten}}]{2006A&A...450..569P}
{Pillai}, T., {Wyrowski}, F., {Carey}, S.~J., \& {Menten}, K.~M.
  2006{\natexlab{a}}, \aap, 450, 569

\bibitem[{{Pillai} {et~al.}(2006{\natexlab{b}}){Pillai}, {Wyrowski}, {Menten},
  \& {Kr{\"u}gel}}]{2006A&A...447..929P}
{Pillai}, T., {Wyrowski}, F., {Menten}, K.~M., \& {Kr{\"u}gel}, E.
  2006{\natexlab{b}}, \aap, 447, 929

\bibitem[{{Pineda} {et~al.}(2009){Pineda}, {Rosolowsky}, \&
  {Goodman}}]{2009ApJ...699L.134P}
{Pineda}, J.~E., {Rosolowsky}, E.~W., \& {Goodman}, A.~A. 2009, \apjl, 699,
  L134

\bibitem[{{Pirogov} {et~al.}(2007){Pirogov}, {Zinchenko}, {Caselli}, \&
  {Johansson}}]{2007A&A...461..523P}
{Pirogov}, L., {Zinchenko}, I., {Caselli}, P., \& {Johansson}, L.~E.~B. 2007,
  \aap, 461, 523

\bibitem[{{Purcell} {et~al.}(2006){Purcell}, {Balasubramanyam}, {Burton},
  {Walsh}, {Minier}, {Hunt-Cunningham}, {Kedziora-Chudczer}, {Longmore},
  {Hill}, {Bains}, {Barnes}, {Busfield}, {Calisse}, {Crighton}, {Curran},
  {Davis}, {Dempsey}, {Derragopian}, {Fulton}, {Hidas}, {Hoare}, {Lee}, {Ladd},
  {Lumsden}, {Moore}, {Murphy}, {Oudmaijer}, {Pracy}, {Rathborne}, {Robertson},
  {Schultz}, {Shobbrook}, {Sparks}, {Storey}, \&
  {Travouillion}}]{2006MNRAS.367..553P}
{Purcell}, C.~R., {Balasubramanyam}, R., {Burton}, M.~G., {et~al.} 2006,
  \mnras, 367, 553

\bibitem[{{Ragan} {et~al.}(2006){Ragan}, {Bergin}, {Plume}, {Gibson}, {Wilner},
  {O'Brien}, \& {Hails}}]{2006ApJS..166..567R}
{Ragan}, S.~E., {Bergin}, E.~A., {Plume}, R., {et~al.} 2006, \apjs, 166, 567

\bibitem[{{Rathborne} {et~al.}(2010){Rathborne}, {Jackson}, {Chambers},
  {Stojimirovic}, {Simon}, {Shipman}, \& {Frieswijk}}]{2010ApJ...715..310R}
{Rathborne}, J.~M., {Jackson}, J.~M., {Chambers}, E.~T., {et~al.} 2010, \apj,
  715, 310

\bibitem[{{Rathborne} {et~al.}(2006){Rathborne}, {Jackson}, \&
  {Simon}}]{2006ApJ...641..389R}
{Rathborne}, J.~M., {Jackson}, J.~M., \& {Simon}, R. 2006, \apj, 641, 389

\bibitem[{{Rathborne} {et~al.}(2008){Rathborne}, {Jackson}, {Zhang}, \&
  {Simon}}]{2008ApJ...689.1141R}
{Rathborne}, J.~M., {Jackson}, J.~M., {Zhang}, Q., \& {Simon}, R. 2008, \apj,
  689, 1141

\bibitem[{{Sakai} {et~al.}(2008){Sakai}, {Sakai}, {Kamegai}, {Hirota},
  {Yamaguchi}, {Shiba}, \& {Yamamoto}}]{2008ApJ...678.1049S}
{Sakai}, T., {Sakai}, N., {Kamegai}, K., {et~al.} 2008, \apj, 678, 1049

\bibitem[{{Schuller} {et~al.}(2009){Schuller}, {Menten}, {Contreras},
  {Wyrowski}, {Schilke}, {Bronfman}, {Henning}, {Walmsley}, {Beuther},
  {Bontemps}, {Cesaroni}, {Deharveng}, {Garay}, {Herpin}, {Lefloch}, {Linz},
  {Mardones}, {Minier}, {Molinari}, {Motte}, {Nyman}, {Reveret}, {Risacher},
  {Russeil}, {Schneider}, {Testi}, {Troost}, {Vasyunina}, {Wienen}, {Zavagno},
  {Kovacs}, {Kreysa}, {Siringo}, \& {Wei{\ss}}}]{2009A&A...504..415S}
{Schuller}, F., {Menten}, K.~M., {Contreras}, Y., {et~al.} 2009, \aap, 504, 415

\bibitem[{{Shirley} {et~al.}(2011){Shirley}, {Huard}, {Pontoppidan}, {Wilner},
  {Stutz}, {Bieging}, \& {Evans}}]{2011ApJ...728..143S}
{Shirley}, Y.~L., {Huard}, T.~L., {Pontoppidan}, K.~M., {et~al.} 2011, \apj,
  728, 143

\bibitem[{{Simon} {et~al.}(2006){Simon}, {Jackson}, {Rathborne}, \&
  {Chambers}}]{2006ApJ...639..227S}
{Simon}, R., {Jackson}, J.~M., {Rathborne}, J.~M., \& {Chambers}, E.~T. 2006,
  \apj, 639, 227

\bibitem[{{Siringo} {et~al.}(2009){Siringo}, {Kreysa}, {Kov{\'a}cs},
  {Schuller}, {Wei{\ss}}, {Esch}, {Gem{\"u}nd}, {Jethava}, {Lundershausen},
  {Colin}, {G{\"u}sten}, {Menten}, {Beelen}, {Bertoldi}, {Beeman}, \&
  {Haller}}]{2009A&A...497..945S}
{Siringo}, G., {Kreysa}, E., {Kov{\'a}cs}, A., {et~al.} 2009, \aap, 497, 945

\bibitem[{{Tafalla} {et~al.}(2004){Tafalla}, {Myers}, {Caselli}, \&
  {Walmsley}}]{2004A&A...416..191T}
{Tafalla}, M., {Myers}, P.~C., {Caselli}, P., \& {Walmsley}, C.~M. 2004, \aap,
  416, 191

\bibitem[{{Tan}(2007)}]{2007IAUS..237..258T}
{Tan}, J.~C. 2007, in IAU Symposium, Vol. 237, IAU Symposium, ed.
  {B.~G.~Elmegreen \& J.~Palous}, 258--264

\bibitem[{{Testi} {et~al.}(1999){Testi}, {Palla}, \&
  {Natta}}]{1999A&A...342..515T}
{Testi}, L., {Palla}, F., \& {Natta}, A. 1999, \aap, 342, 515

\bibitem[{{Testi} {et~al.}(1997){Testi}, {Palla}, {Prusti}, {Natta}, \&
  {Maltagliati}}]{1997A&A...320..159T}
{Testi}, L., {Palla}, F., {Prusti}, T., {Natta}, A., \& {Maltagliati}, S. 1997,
  \aap, 320, 159

\bibitem[{{Testi} \& {Sargent}(1998)}]{1998ApJ...508L..91T}
{Testi}, L. \& {Sargent}, A.~I. 1998, \apjl, 508, L91

\bibitem[{{Testi} {et~al.}(2000){Testi}, {Sargent}, {Olmi}, \&
  {Onello}}]{2000ApJ...540L..53T}
{Testi}, L., {Sargent}, A.~I., {Olmi}, L., \& {Onello}, J.~S. 2000, \apjl, 540,
  L53

\bibitem[{{Williams} {et~al.}(2000){Williams}, {Blitz}, \&
  {McKee}}]{2000prpl.conf...97W}
{Williams}, J.~P., {Blitz}, L., \& {McKee}, C.~F. 2000, Protostars and Planets
  IV, 97

\bibitem[{{Williams} {et~al.}(1994){Williams}, {de Geus}, \&
  {Blitz}}]{1994ApJ...428..693W}
{Williams}, J.~P., {de Geus}, E.~J., \& {Blitz}, L. 1994, \apj, 428, 693

\bibitem[{{Wilson} \& {Rood}(1994)}]{1994ARA&A..32..191W}
{Wilson}, T.~L. \& {Rood}, R. 1994, \araa, 32, 191

\bibitem[{{Zhang} {et~al.}(1999){Zhang}, {Hunter}, {Sridharan}, \&
  {Cesaroni}}]{1999ApJ...527L.117Z}
{Zhang}, Q., {Hunter}, T.~R., {Sridharan}, T.~K., \& {Cesaroni}, R. 1999,
  \apjl, 527, L117

\end{thebibliography}
\end{document}